\documentclass[twocolumn]{aastex631}

\usepackage{amsmath}	
\usepackage{hyperref}
\usepackage{xcolor}

\newcommand{\src}{MXB~1659-29}
\newcommand{\saxj}{SAX~J1808.4-3658}
\newcommand{\durca}{dUrca}

\submitjournal{ApJ}

\shorttitle{Fast neutrino cooling in MXB 1659-29}
\shortauthors{Mendes et al.}

\begin{document}

\title{Fast neutrino cooling in the accreting neutron star MXB 1659-29}

\correspondingauthor{Melissa Mendes}
\email{melissa.mendessilva@mail.mcgill.ca}

\author[0000-0002-5250-0723]{Melissa Mendes}
\affiliation{Department of Physics and McGill Space Institute, McGill University, 3600 rue University, Montreal, QC, H3A 2T8, Canada}

\author[0000-0002-7371-3656]{Farrukh J. Fattoyev}
\affiliation{Department of Physics, Manhattan College, Riverdale, New York 10471, USA}

\author[0000-0002-6335-0169]{Andrew Cumming}
\affiliation{Department of Physics and McGill Space Institute, McGill University, 3600 rue University, Montreal, QC, H3A 2T8, Canada}

\author[0000-0001-7351-9338]{Charles Gale}
\affiliation{Department of Physics, McGill University, 3600 rue University, Montreal, QC, H3A 2T8, Canada}

\begin{abstract}  
Modelling of crust heating and cooling across multiple accretion outbursts of the low mass X-ray binary \src\ indicates that the neutrino luminosity of the neutron star core is consistent with direct Urca reactions occurring in $\sim 1$\% of the core volume. We investigate this scenario with neutron star models that include a detailed equation of state parametrized by the slope of the nuclear symmetry energy $L$, and a range of neutron and proton superfluid gaps. We find that the predicted neutron star mass depends sensitively on $L$ and the assumed gaps. We discuss which combinations of superfluid gaps reproduce the inferred neutrino luminosity. Larger values of $L\gtrsim 80\ {\rm MeV}$ require superfluidity to suppress \durca\ reactions in low mass neutron stars, i.e. that the proton or neutron gap is sufficiently strong and extends to high enough density. However, the largest gaps give masses near the maximum mass, making it difficult to accommodate colder neutron stars. The heat capacities of our models span the range from fully-paired to fully-unpaired nucleons meaning that long term observations of core cooling could distinguish between models. As a route to solutions with a larger emitting volume, which could provide a more natural explanation for the inferred neutrino luminosity, we discuss the possibility of alternative, less efficient, fast cooling processes in exotic cores. To be consistent with the inferred neutrino luminosity, such processes must be within a factor of $\sim 1000$ of \durca.
We discuss the impact of future constraints on neutron star mass, radius and the density dependence of the symmetry energy.  
\end{abstract}


\section{Introduction}

Neutron stars in transiently-accreting low mass X-ray binaries (LMXBs) are remarkable laboratories to probe the physics of dense matter (for a review see \citealt{Wijnands2017}). While accreting, the neutron star crust is heated by accretion-induced nuclear reactions, with most of the energy flowing inwards to the neutron star core. After accretion ends, in the quiescent phase, the neutron star surface temperature can be measured, which in turn gives an estimation of the neutron star core temperature. The quiescent temperatures and luminosities of LMXB neutron stars have been used to infer the efficiency of neutrino emission processes and superfluid state in their cores \citep{Yakovlev2004ARAA,Heinke2007,Levenfish2007,Heinke2009,Wijnands2013,Beznogov2015a,Beznogov2015b,Han2017,Potekhin2019}, the thermal conductivity and superfluidity of the neutron star crust \citep{Shternin2007,Brown2009,PageReddy2012}, and the heat capacity of the core \citep{Brownetal2017,Degenaar2021}.

There is growing evidence that a number of neutron stars have highly-efficient fast neutrino processes in their cores, based on very low quiescent temperatures \citep{Heinke2007,Heinke2009,Han2017, Potekhin2019}.
Characterized by a local emissivity $\propto T^6$, where $T$ is the local temperature, the most efficient fast neutrino process is the direct Urca (\durca) process in nucleonic matter in which neutrinos are produced by the reactions \citep{Lattimer1991}
\begin{equation}\label{eq:\durca_reactions}
n \rightarrow p+e^{-}+\bar{\nu}_{e}, \quad p+e^{-} \rightarrow n+\nu_{e}.
\end{equation}
Momentum conservation in these reactions means that they can proceed only if the proton fraction is sufficiently large ($Y_p\gtrsim 1/9$). This means that determining that the \durca\ process is happening in a neutron star core directly constrains the proton fraction and therefore the value of the nuclear symmetry energy at high density (see discussion in \citealt{Lattimer2018}). It also means that the central density and therefore mass of the neutron star is large enough to achieve the critical value of $Y_p$. With more exotic compositions, such as a meson condensate or quark matter, other fast processes are possible, with the same $\propto T^6$ scaling but a typically smaller normalization (see \citealt{Yakovlev2001} for a review). The core temperature is therefore a very interesting quantity that can be constrained by observations and depends on the unknown composition of neutron star cores.

The LMXB \src\ has shown multiple accretion outbursts in which the neutron star crust has been observed to thermally-relax in quiescence \citep{Parikh2019}. This is unusual because most LMXBs with multiple outbursts have short, frequent outbursts that do not significantly heat the crust at depth; while the LMXBs with large outbursts that heat the crust significantly, have typically only shown one outburst because the recurrence time between outbursts is very long. \src\ went into outburst in the late 1970s, in 2001 and again in 2015, with each outburst lasting $\sim 2$ years \citep{Cackett2006,Parikh2019}. 

By modelling the sequence of outbursts in \src\ and the relaxation of the surface temperature in quiescence, \cite{Brownetal2018} showed that the core must be cooled by a fast neutrino process. The rate at which the neutron star cools after each outburst depends on the temperature of the neutron star crust, set by the rate at which the neutron star core is being heated and therefore cooled by neutrinos. The composition of the neutron star envelope is also constrained by the shape of the cooling curve, reducing an uncertainty in mapping the surface temperature to core temperature \citep{Brownetal2017} (for \src, the inferred core temperature is $\approx 2.5\times 10^7\ {\rm K}$). Slow neutrino processes, \textit{i.e}.~less efficient processes such as modified URCA with emissivity $\propto T^8$, cannot provide the required neutrino luminosity at this core temperature.

Computing an average value of $L_\nu/T^6$ for the core and comparing with the expected value for \durca, \citep{Brownetal2018} found that the neutrino cooling luminosity of \src\ is consistent with \durca\ occurring over about $\sim 1$\% of the core volume. 
Such a low effective emitting volume gives an interesting constraint on the neutron star in \src. Possible explanations are that (1) the neutron star has a mass within a few percent of the mass at which \durca\ reactions become possible, (2) superfluidity suppresses the \durca\ reactions throughout most of the available volume, reducing the overall luminosity, or (3) a less efficient fast process is operating over the core. \cite{Brownetal2018} pointed out that these possibilities could in principle be distinguished because they make different predictions for the cooling rate of the core in quiescence because the heat capacity of the core depends on its composition and the extent of superfluidity.

In this paper, we explore the different scenarios for the neutrino emission of \src\ using detailed neutron star models that include a variety of superfluid gap models. We use an equation of state parametrized by the slope of the nuclear symmetry energy $L$, since this parameter determines the proton fraction at high density and therefore the onset density for \durca\ reactions. We describe the input microphysics that we use in \S \ref{sec:formalism}. In \S \ref{sec:results}, we investigate the values of $L$ and neutron star mass that are required to reproduce the inferred neutrino luminosity of \src\ under different assumptions about the superfluid gap. In \S \ref{sec:heatcapacity}, we calculate the heat capacity of our models since this is a potential observable that can distinguish the different scenarios. We end with a discussion of our results in \S \ref{sec:discussion}, including the possibility of less efficient fast emission processes that might occur if exotic particles are present in the core, and potential future observational and experimental constraints.

\section{Details of the calculation and input microphysics}
\label{sec:formalism}

\subsection{Equation of state and neutron star structure}

The family of equations of state (EOS) we use to describe the core of neutron stars is based on relativistic mean field (RMF) model FSUGold2 \citep{FSUGoldReference}. This EOS was one of the first to reproduce not only ground-state properties of finite nuclei, but also the maximum observed neutron star mass at the time. A detailed framework of the EOS we generate is available in \cite{Mendes2021}, but we summarize their main characteristics here, for convenience.

\begin{figure}
\centering
\includegraphics[width=0.98\columnwidth]{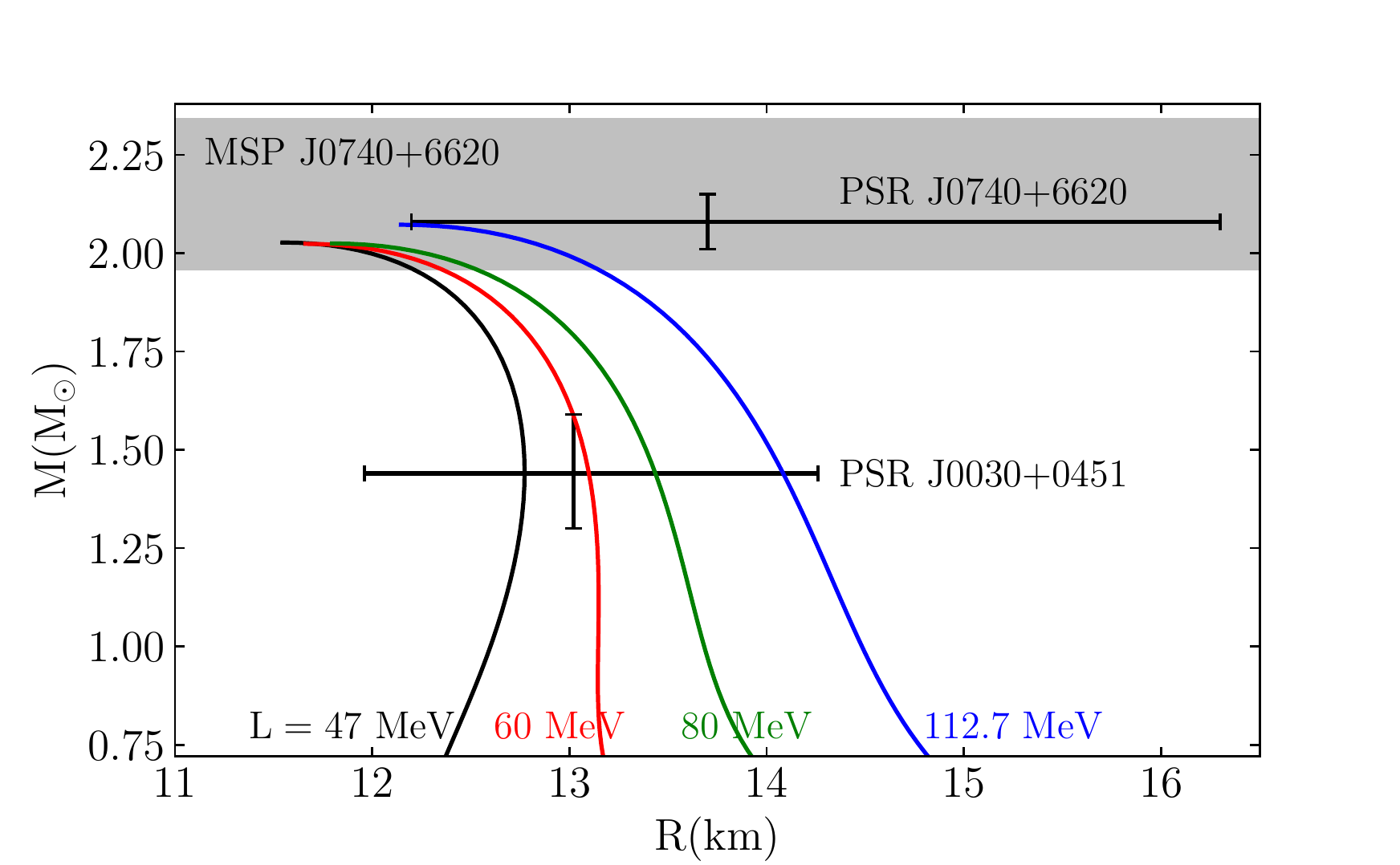}
\caption{Mass versus radius curves of equations of state with $L=47$ MeV, $L=60$ MeV, $L=80$ MeV and $L=112.7$ MeV, from left to right. Observational constraints for $\mathrm{PSR} \, J0030+0451$ \citep{Miller:2019cac} and $\mathrm{PSR} \, J0740+6620$ \citep{Miller_2021} indicated with error bars and maximum mass neutron star observed to date, $\mathrm{MSP} \, J0740+6620$ \citep{NANOGrav:2019jur}, with shaded region.}
\label{fig:mxr}
\end{figure}

We consider a minimal model in which only neutrons, protons, electrons and muons are the particle constituents of neutron stars. Consider the expansion of the total energy per nucleon $E(\rho, \alpha)$ at zero temperature,
\begin{equation}
E(\rho, \alpha) = E_{\rm SNM}(\rho) + E_{\text {\rm sym}}(\rho) \cdot \alpha^{2}+\mathcal{O}\left(\alpha^{4}\right) 
\label{EnNucleon} \,
\end{equation}
where $\rho = \rho_{\rm n} + \rho_{\rm p}$ is the total baryon number density and $\alpha=(\rho_{\rm n} - \rho_{\rm p})/\rho$ is the neutron-proton asymmetry parameter. Next, consider the Taylor series \citep{Piekarewicz:2008nh} that characterize both the energy per nucleon in symmetric nuclear matter (SNM), $E_{\rm SNM}(\rho)$, and the symmetry energy, $E_{\text {\rm sym}}(\rho)$, near the nuclear saturation density $\rho_{\rm sat} = 0.15$ fm$^{-3}$,
\begin{equation} 
\begin{split}
& E_{\rm SNM}(\rho) = B + \frac{1}{2} K x^2 + \cdots 
\,\\
& E_{\rm sym}(\rho) = J + L x+ \frac{1}{2} K_{\rm sym} x^2 + \cdots\,\\
& \mathrm{where}\, x = (\rho - \rho_{\rm sat})/3\rho_{\rm sat}.
\end{split}
\end{equation}
The family of the EOS we work with shares identical SNM bulk parameters, such as the energy per nucleon $B=-16.26$ MeV and incompressibility coefficient $K=237.7$ MeV. The symmetry energy $\tilde{J}$ at a subsaturation density of $\rho =0.1$ fm$^{-3}$ is also fixed to ensure that binding energies and charge radii of finite nuclei are well reproduced. 
Their different slope of symmetry energy $(L)$, varying from $47$ MeV to $112.7$ MeV, provide distinct neutron skin thicknesses and neutron star properties, such as radii, all consistent with the current experimental and observational data \citep{Adhikari:2021phr, Riley:2019yda, Miller:2019cac, Abbott:PRL2017, Abbott:2018exr}. 
In particular, increasing $L$ leads to a larger symmetry energy at supersaturation densities, which increases the proton fraction $Y_{\rm p} = \rho_{\rm p}/\rho$ in the innermost region of the star. 
In addition, increasing $L$ leads to larger neutron star radii. The mass-radius curves for different $L$ values are shown in Figure~\ref{fig:mxr}.

The outer crust is described by the EOS from \cite{Baym1971} and the inner crust, by the EOS from \cite{Negele1973}. We assume a non-rotating spherically-symmetric neutron star, and solve the Tolman-Oppenheimer-Volkoff (TOV) equations
\begin{equation}\label{TOV}
\begin{split}
\frac{\mathrm{d} P}{\mathrm{d} r} &=-\frac{\mathcal{E}(r)}{c^2}\frac{G m(r)}{r^2} \left[1 + \frac{P(r)}{\mathcal{E}(r)}\right] \\
& \qquad \quad \left[1+\frac{4 \pi r^{3} P(r)}{m(r) c^2}\right] \left[1-\frac{2Gm(r)}{c^2r}\right]^{-1} \\
\frac{\mathrm{d} m}{\mathrm{d} r} &= 4 \pi r^2 \frac{\mathcal{E}(r)}{c^2}\\
\frac{\mathrm{d} \phi}{\mathrm{d} r} &=- \frac{1}{\mathcal{E}(r)+P(r)} \frac{\mathrm{d} P}{\mathrm{d} r}, 
\end{split}
\end{equation}
where $m(r)$ is the mass within radius $r$, $P(r)$ is the pressure, $\mathcal{E}(r)$ is the energy density and $\phi(r)$ is the gravitational potential such that at the surface of the star, $r = R$ and $m=M$, the pressure vanishes, $P(R) = 0$ and $\phi(R)=\frac{1}{2} \ln (1-{2 G M}/{c^2R})$, $G$ is the gravitational constant.

\subsection{Neutrino emissivity}

Since our goal is to reproduce the inferred neutrino luminosity of \src, we consider the fast cooling process of \durca\ only. If there are no muons participating, \durca\ cooling takes place through the reactions in equation (\ref{eq:\durca_reactions}) which conserve momentum only if
\begin{equation}\label{eq:triangle}
k_{F n} \leq k_{F p}+k_{F e},
\end{equation}
which implies that for \durca\ reactions the proton fraction must exceed a threshold value 
\begin{equation}\label{condition}
Y_{p} \geq Y_{\mathrm{p\, \durca}} = \left[Y_{n}^{1/3}-Y_{e}^{1/3}\right]^{3},
\end{equation} 
as explained in \cite{Yakovlev2001}. Here, $k_{\mathrm{Fx}}$ are the Fermi momenta, a function of $\rho_\mathrm{x}$, the number density for each species and the particle fraction $Y_{\mathrm{x}}$. When muons participate, additional \durca\ reactions take place 
\begin{equation}\label{muonsincluded}
    n \rightarrow p+\mu^{-}+\bar{\nu}_{\mu^{-}}, \quad p+\mu^{-} \rightarrow n+\nu_{\mu^{-}}
\end{equation}
which has its own threshold given by replacing $k_{F e}$ with $k_{F \mu}$ in equation (\ref{eq:triangle}). As well as introducing an additional neutrino producing reaction, muons also modify the electron fraction $Y_e = Y_p - Y_\mu$ which enters equation (\ref{condition}), and so modify the electron \durca\ channel even before the threshold for muon \durca\ is reached.

The threshold proton fraction corresponds to a threshold density $\rho_\mathrm{\durca}$ for \durca\ processes to occur. Only in regions of the core with $\rho>\rho_\mathrm{\durca}$ is \durca\ allowed, and this also implies that only neutron stars massive enough to have a central density $\rho_c>\rho_\mathrm{\durca}$ can cool by \durca.
The \durca\ neutrino luminosity, as seen by an observer at infinity, is given by
\begin{equation}\label{integration}
L_{\nu_{\mathrm{\durca}}}^{\infty} = \int_0^{R_\mathrm{core}}\frac{4 \pi r^2 \epsilon_{0}^{\mathrm{\durca}, \mathrm{total}} e^{2 \, \phi (r)}}{\left(1-2 G m(r)/c^ 2 r\right)^{1/2}}\, dr,
\end{equation}
where the integral is over the neutron star core and the local neutrino emissivity is \citep{Yakovlev2001} 
\begin{equation} \label{emissivity}
\begin{split}
    \epsilon_{0}^{\mathrm{\durca}, e^{-}} &=\frac{457 \pi}{10080} G_{\mathrm{F}}^{2} \cos ^{2} \theta_{\mathrm{C}}\left(1+3 g_{\mathrm{A}}^{2}\right)\\
    & \qquad \times \frac{m_{n}^{*} m_{\mathrm{p}}^{*} m_{e}}{h^{10} c^{3}}\left(k_{\mathrm{B}} T\right)^{6} \Theta_{\mathrm{npe}}\\
    \epsilon_{0}^{\mathrm{\durca}, \mu^{-}} &= \epsilon_{0}^{\mathrm{\durca},e^{-}} \Theta_{\mathrm{np\mu}}\\
    \epsilon_{0}^{\mathrm{\durca}, \mathrm{total}} &= \epsilon_{0}^{\mathrm{\durca}, e^{-}}+ \epsilon_{0}^{\mathrm{\durca},\mu^{-}},\\
\end{split}
\end{equation}
where we use the weak coupling constant $G_{\mathrm{F}}= 1.436\times 10^{-62}\, \mathrm{J} \mathrm{m}^3 $ and Cabibbo angle $\sin \theta_{\mathrm{C}} = 0.228$ \citep{Zyla2020PDG}, and in-medium axial vector coupling constant from \cite{Carter:2002}, $g_{\mathrm{A}}=-1.2601 (1-\rho/(4.15 (\rho_{0}+\rho))$. We account for in-medium interactions through $\mathrm{m}_{x}^{*}$, which represents the Landau effective mass of species $x$ defined as $\mathrm{m}^{*}=\sqrt{m_{\rm D}^2+(\hbar k_{\rm F}/c)^2}$ with $m_{\rm D}$ being the nuclear interaction-dependent Dirac effective mass (see, e.g. \citealt{Chen:2007ih}) and $k_{\rm F}$ is the Fermi wave-number of the nucleon. $\Theta_{\mathrm{npe(\mu)}}$ is a step function that restricts direct Urca reactions to the regions with $\rho>\rho_\mathrm{\durca}$.

\subsection{Treatment of superfluidity and superconductivity}
\label{sec:intro_superfluidity}

Including superfluidity and superconductivity in the neutron star core model changes the neutrino luminosity, since the local neutrino emissivity is exponentially reduced by a reduction factor $R_L$, giving $\epsilon^{\mathrm{\durca}}=\epsilon_{0}^{\mathrm{\durca}} R_L$ \citep{Yakovlev2001}. We consider both proton singlet (PS) (${ }^{1} \mathrm{S}_{0}$) and neutron triplet (NT) $({ }^{3} \mathrm{P}_{2},\,m_{J}=0)$ pairing in the core. For proton singlets, the reduction factor is given by \citep{Yakovlev2001}
\begin{equation} \label{reduced1}
\begin{split}
    R_{\mathrm{L}} &=\left[0.2312+\sqrt{(0.7688)^{2}+(0.1438 ~ v_{\mathrm{S}})^{2}}\right]^{5.5}\\
    & \qquad \times \exp \left(3.427-\sqrt{(3.427)^{2}+v_{\mathrm{S}}^{2}}\right),
    \\
    v_{\mathrm{S}} &=\sqrt{1-\tau}\left(1.456-\frac{0.157}{\sqrt{\tau}}+\frac{1.764}{\tau}\right),
\end{split}
\end{equation}
while for neutron triplets,
\begin{equation} \label{reduced2}
\begin{split}
    R_{\mathrm{L}} &=\left[0.2546+\sqrt{(0.7454)^{2}+(0.1284 \, v_{\mathrm{T}})^{2}}\right]^{5}\\
    & \qquad \times \exp \left(2.701-\sqrt{(2.701)^{2}+v_{\mathrm{T}}^{2}}\right),\\
    v_{\mathrm{T}} &=\sqrt{1-\tau}\left(0.7893+\frac{1.188}{\tau}\right).
\end{split}
\end{equation}
Here $\tau = T/T_c$, where $T_c$ is the critical temperature, calculated according to each gap model parametrization, and $T$ is the local temperature at radius $r$, $T(r) = \tilde{T} \exp(-\phi(r))$, with $\tilde{T}$ the temperature of the isothermal core as measured at infinity. 

When proton singlet superconductivity and neutron triplet superfluidity are simultaneously active, we use the approximation
\begin{equation}
R_{L} \sim \min \left(R_{L,\mathrm{singlet}}, R_{L,\mathrm{triplet}}\right)
\end{equation}
which is valid in the limit of strong superfluidity \citep{Yakovlev1994}. A more accurate calculation could be performed with combinations of the asymptotic expressions described in \cite{Yakovlev1994}, however, since $T_c \gg T$ except in a narrow range of density, they would only provide minor corrections.   

The heat capacity of neutrons and protons is similarly reduced when they are superfluid or superconducting, $\mathrm{C}_{p,n}^{\mathrm{superfluid}} = \mathrm{C}_{p,n}\,R_C$ \citep{Yakovlev:1994}, where, for proton singlets, 
\begin{equation} \label{reduced3}
\begin{split}
    R_C &=\left[0.4186+\sqrt{(1.007)^{2}+(0.5010 \, u_{\mathrm{S}})^{2}}\right]^{2.5} \\
    & \qquad \times \exp \left(1.456-\sqrt{(1.456)^{2}+u_{\mathrm{S}}^{2}}\right),\\
    u_{\mathrm{S}} &=\sqrt{1-\tau}(1.456-0.157 / \sqrt{\tau}+1.764 / \tau),
\end{split}
\end{equation}
and for neutron triplets,
\begin{equation} \label{reduced4}
\begin{split}
    R_{C} &= \left[0.6893+\sqrt{(0.790)^{2}+(0.03983 \, u_{\mathrm{T}})^{2}}\right]^{2} \\
    & \qquad \times \exp \left(1.934-\sqrt{(1.934)^{2}+\frac{u_{\mathrm{T}}^{2}}{16 \pi}}\right),\\
    u_{\mathrm{T}} &=\sqrt{1-\tau}(5.596+8.424 / \tau).
\end{split}
\end{equation}
The total heat capacity is given by
\begin{equation} \label{heatcap}
    C^{\mathrm{core}}_{\mathrm{total}} = \int_0^{R_\mathrm{core}}\frac{4 \pi r^2 \sum C_{x}}{\left(1-2 G m(r)/c^ 2 r\right)^{1/2}}\, dr,
\end{equation}
where $C_{x}$  is the contribution to the local heat capacity from each particle species \citep{Yakovlev:1994},
\begin{equation}
    C_x=\frac{m_x^*p_{F, x}}{3 \hbar^{3}} k_{B}^{2} T.
\end{equation}

\begin{figure}
\centering
\includegraphics[width=0.98\columnwidth]{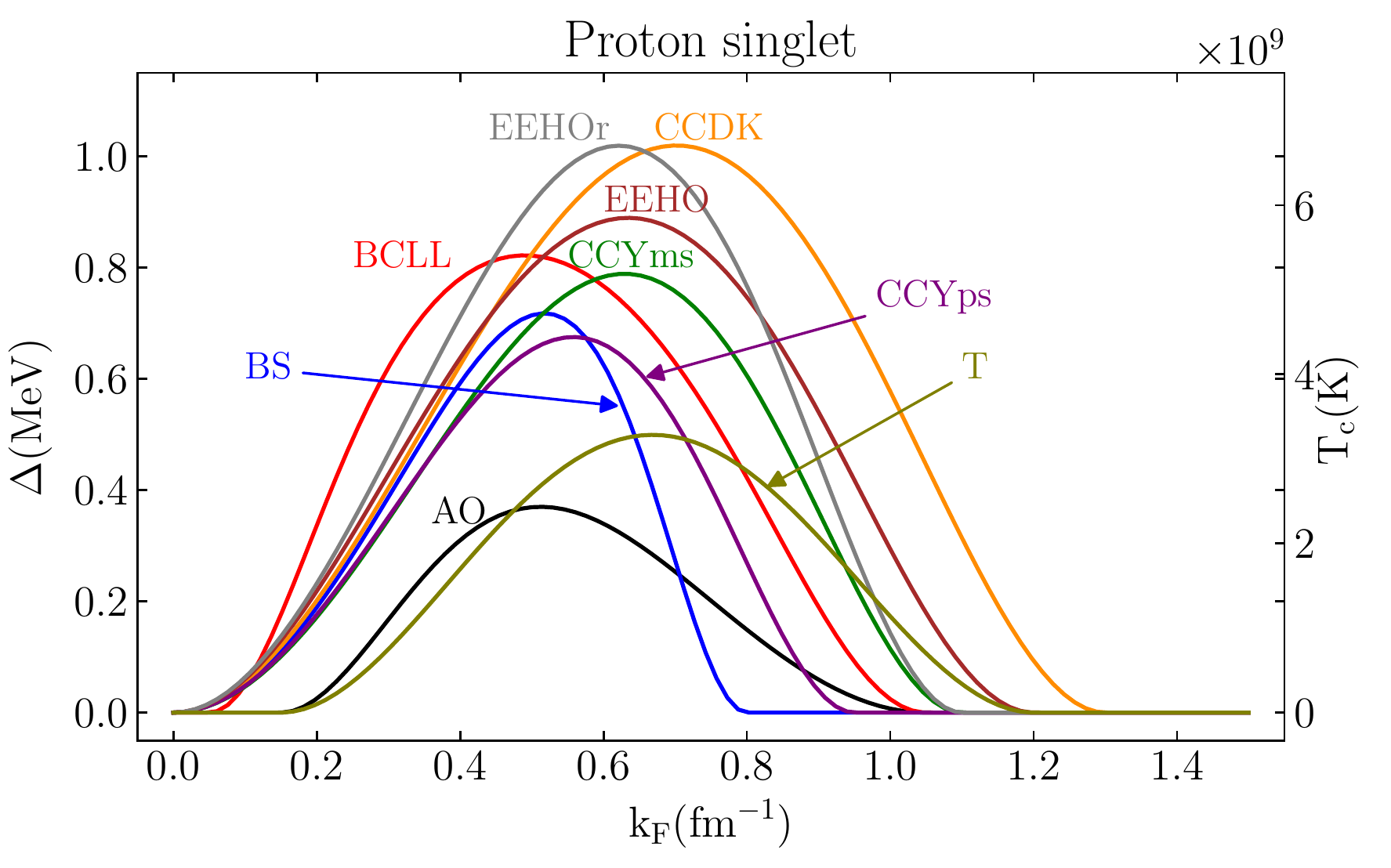}
\includegraphics[width=\columnwidth]{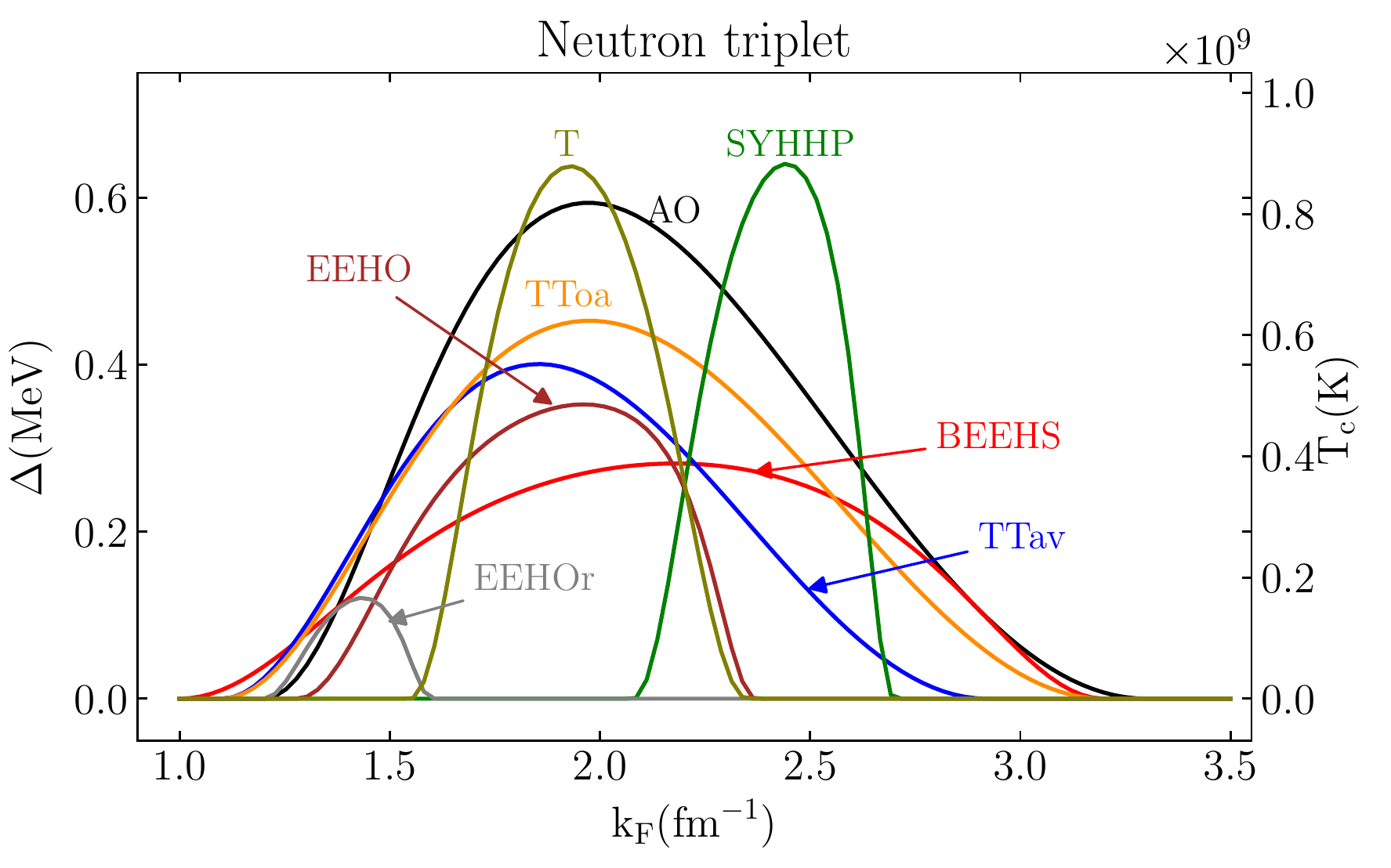}
\caption{Superfluid gap $\Delta \, (\mathrm{MeV})$ and critical temperature $\mathrm{T}_\mathrm{c} \, (\mathrm{K})$ as a function of Fermi momentum $k_F \, (\mathrm{fm^{-1}})$ for proton singlet (top panel) and neutron triplet (bottom panel) gap models used in our calculations (see \citealt{Hoetal2015} for details of the analytic fits and references for each gap model).
}
\label{fig:parametrizations}
\end{figure}

\begin{figure*}
\centering
\includegraphics[width=0.3\textwidth]{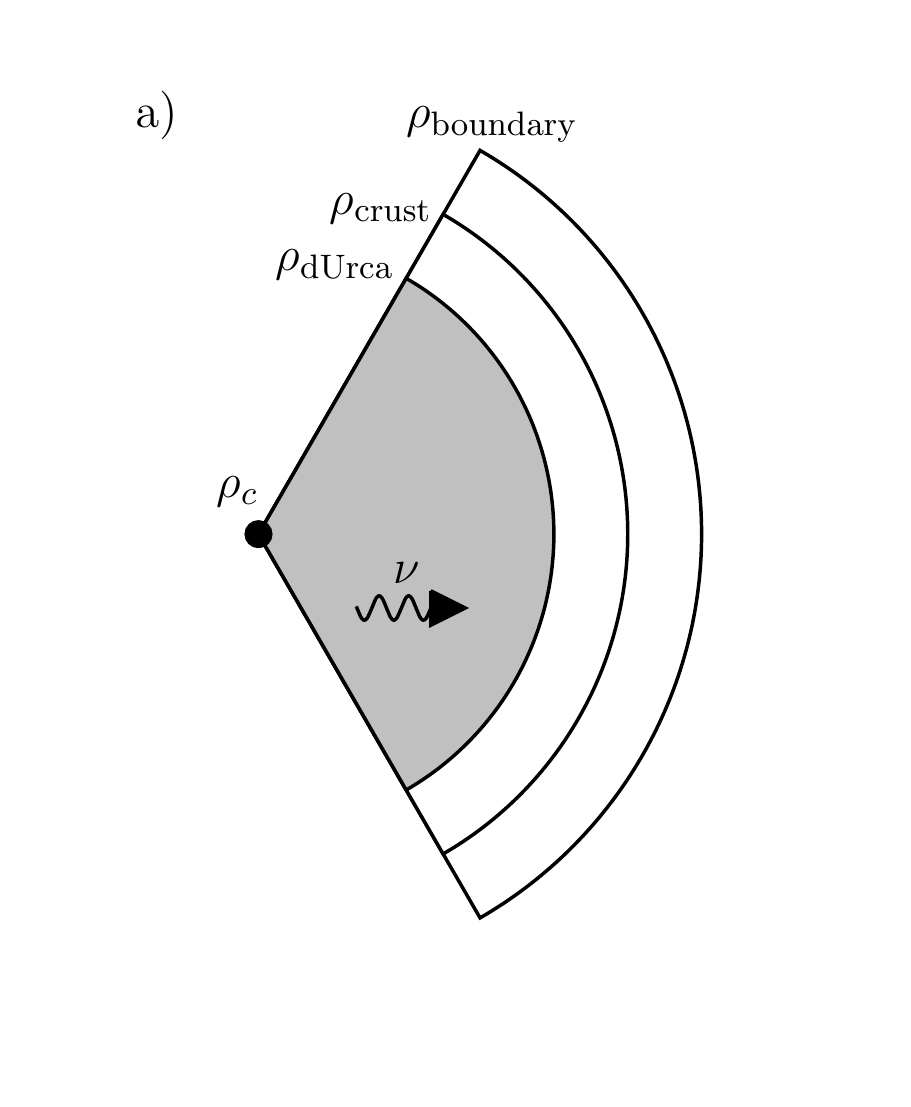}
\includegraphics[width=0.3\textwidth]{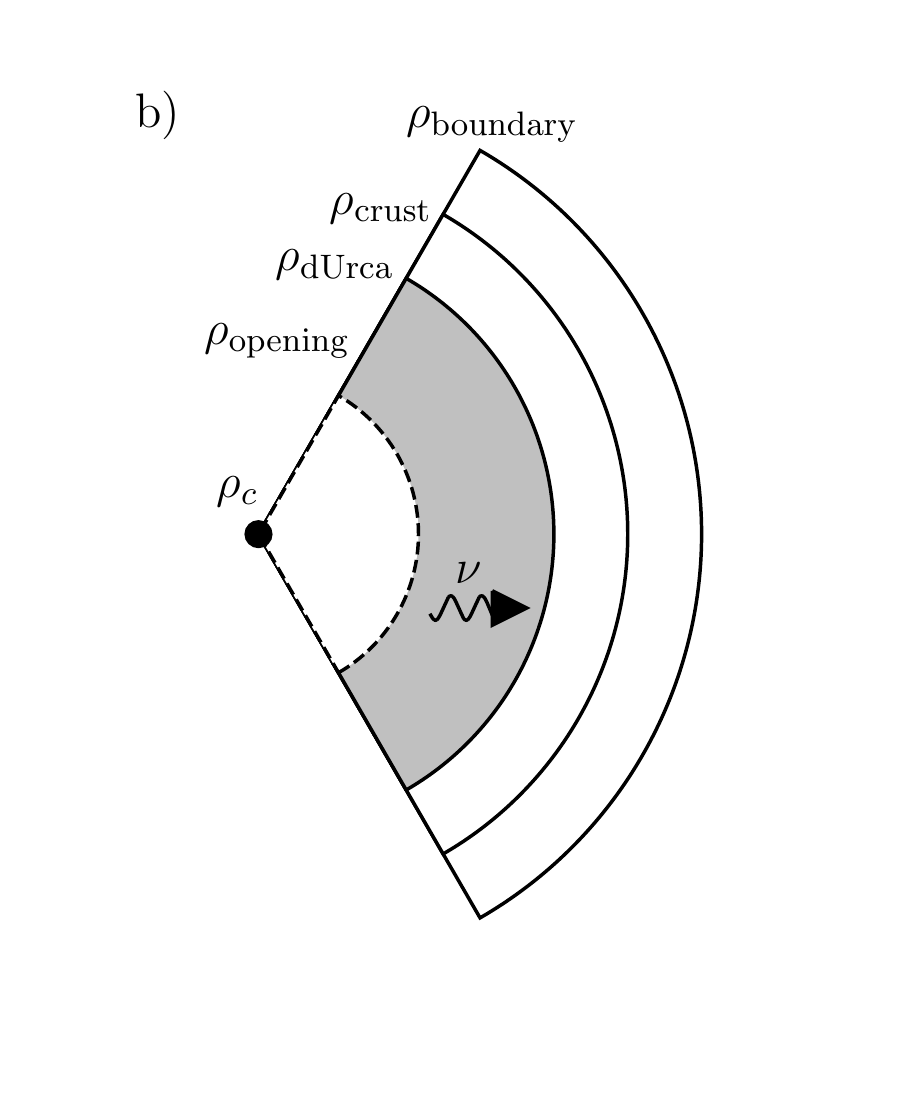}
\includegraphics[width=0.3\textwidth]{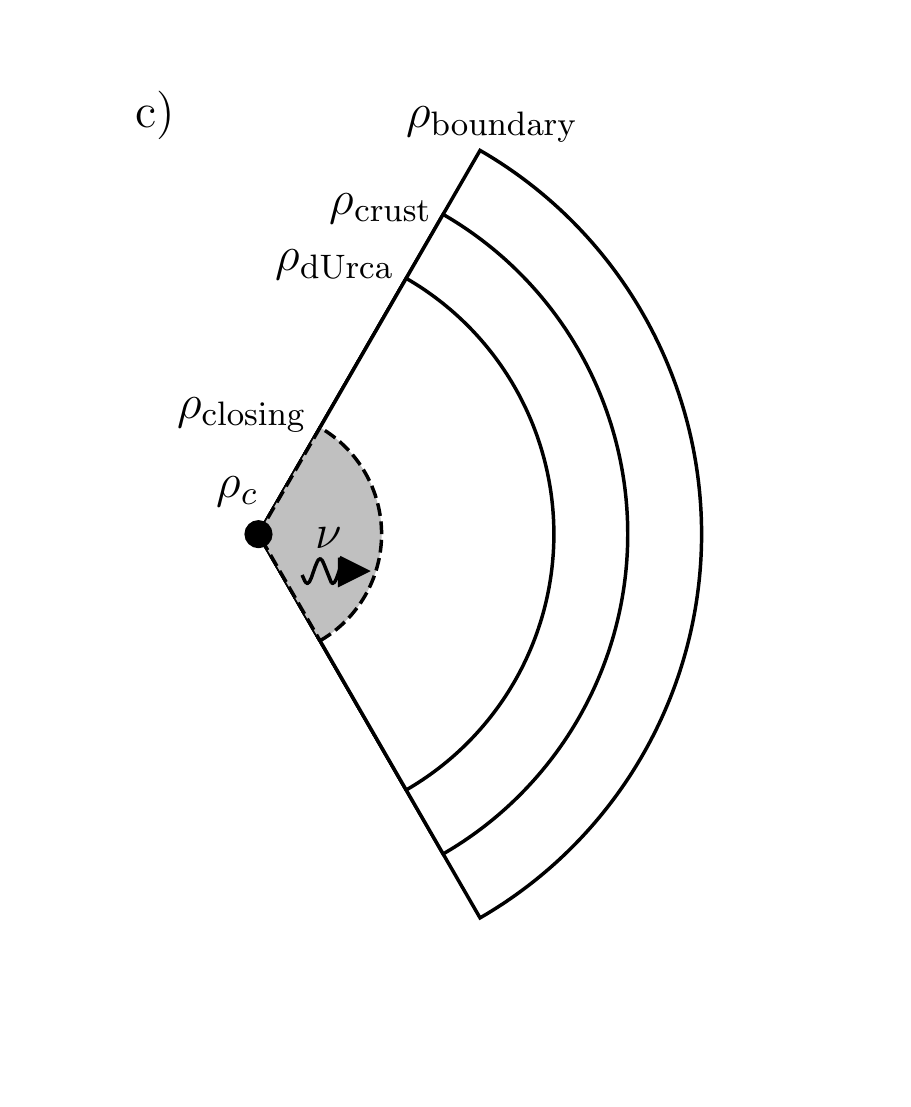}
\caption{Three scenarios to explain the neutrino luminosity of \src. The shaded region indicates the part of the neutron star core undergoing active \durca\ neutrino emission, indicated by the arrow.  (a) No superfluidity in the region $\rho>\rho_\mathrm{\durca}$. (b) A superfluid gap opens at a density $\rho_{\rm opening}>\rho_\mathrm{\durca}$, suppressing \durca\ in the centre and leading to neutrino emission from a shell. (c) The superfluid gap opens before the onset of \durca, $\rho_{\mathrm{opening}} < \rho_{\mathrm{\durca}}$, but closes again at high density $\rho_{\rm closing}<\rho_c$, where $\rho_c$ is the central density. In this case neutrino emission is suppressed in the outer part of the core, reducing the size of the central emitting volume.
Note that distances are not to scale in these diagrams; $\rho_\mathrm{boundary}$, $\rho_\mathrm{crust}$, and $\rho_c$ represent the surface, crust-core boundary and central densities respectively. An example of case b (case c) is the $1.6 \, \mathrm{M}_{\odot}$, $L=70$ MeV  ($2.01 \, \mathrm{M}_{\odot}$, $L=50$ MeV) star with the $\mathrm{NT} \, \mathrm{SYHHP}$ gap shown in Fig.~\ref{fig:newR}.
	\label{illustration}}
\end{figure*}

\subsection{Gap models}

To explore a range of different superfluid gap models, we use the analytic fits of \cite{Hoetal2015} (see their eq.~[2] and Table~II) to nine proton singlet (PS) and eight neutron triplet (NT) gap models. 
The PS gap models are AO \citep{Amundsen1985singlet}, BCLL \citep{Baldo1992}, BS \citep{Baldo2007}, CCDK \citep{Chen1993}, CCYms/CCYps \citep{Chao1972}, EEHO \citep{Elgaroy1996b}, EEHOr \citep{Elgaroy1996a}, and T \citep{Takatsuka1973}.
The NT gap models are AO \citep{Amundsen1985triplet}, BEEHS \citep{Baldo1998}, EEHO \citep{Elgaroy1996NT}, EEHOr \citep{Elgaroy1996a}, SYHHP \citep{Shternin2011}, T \citep{Takatsuka1972}, and TTav/TToa \citep{Takatsuka2004}. For additional references and details of the fits, see \cite{Hoetal2015}.
Unless it is clear from the context, we will prefix the gap name by either NT or PS to make clear whether we are referring to a neutron triplet or proton singlet model, e.g.~NT~AO refers to the neutron triplet AO model. 

Figure \ref{fig:parametrizations} shows these gap models as a function of $k_F$. To help characterize the region of the star which is superfluid, we define the opening $\rho_\mathrm{opening}$ and closing $\rho_{\rm closing}$ densities to correspond to the densities where the local temperature equals the critical temperature, $T_c=T(r)$. The opening and closing densities depend on the EOS (which maps $k_F$ to density for each species), so we compute them for each neutron star model. Suppression of the \durca\ emissivity will occur in the density range $\rho_{\rm opening}\lesssim\rho\lesssim\rho_{\rm closing}$. Our list of gap models covers a range of amplitudes and widths of the critical temperatures for nuclear pairings in neutron star cores, as well as early (low density) and late (high density) openings and closings, and so will allow us to explore the range of expected behavior.

\section{Models of \src\ with dUrca neutrino cooling}
\label{sec:results}

In this section, we attempt to reproduce the inferred neutrino luminosity of \src\ with neutron star models in which the neutrino emission is by the nucleonic \durca\ process, and considering different gap models for neutron and proton superfluidity. We start by considering models without superfluidity (\S \ref{sec:noSF}), then consider the effect of neutron and proton pairing separately (\S \ref{sec:SF}) and in combination (\S \ref{PsNt}). 

Since the mass of the neutron star in \src\ is unconstrained, we take the approach of calculating the range of allowed masses that are consistent with the inferred neutrino luminosity $L_\nu^\infty$ of \src. We take the central value inferred by \cite{Brownetal2018}, $L_{\nu}^{\infty} = 3.9 \times 10^{34} \, \mathrm{erg}/\mathrm{s}$, and also consider upper and lower values $L_{\nu}^{\infty} = 2 \times 10^{34} \, \mathrm{erg}/\mathrm{s}$ and $L_{\nu}^{\infty} = 7.8 \times 10^{34} \, \mathrm{erg}/\mathrm{s}$ which correspond approximately to the $1$-$\sigma$ range found by \cite{Brownetal2018} (see their Fig.~2). We also set the core temperature at infinity to $\tilde{T}=2.5\times 10^7\ {\rm K}$ \citep{Brownetal2018} (note that $\tilde{T}$ is independent of radial coordinate in an isothermal star).

\subsection{Models with no pairing}\label{sec:noSF}

We first consider models without nuclear pairing, so that neutrino cooling occurs from all parts of the neutron star core where the density exceeds the \durca\ threshold density. This situation is indicated  schematically in Fig.~\ref{illustration}a.

\begin{figure}
    \centering
    \includegraphics[width=0.51\textwidth]{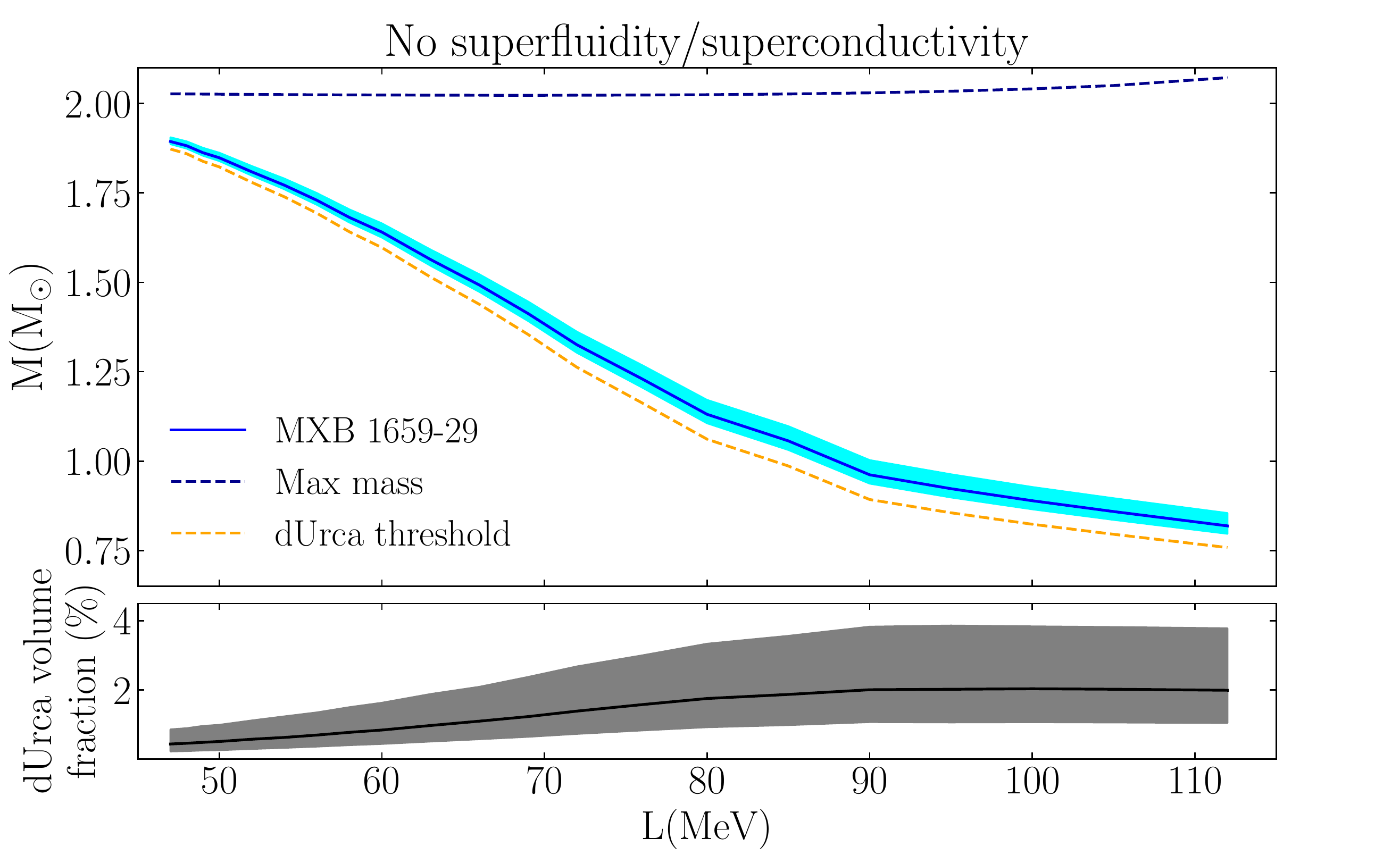}
    \caption{{\em Top panel: } Neutron star mass that reproduces the neutrino luminosity of \src\ as a function of $L$ for neutron star models without pairing. The blue shaded region corresponds to the $1\sigma$ uncertainty in neutrino luminosity from \citealt{Brownetal2018}. The \durca\ threshold and maximum neutron star mass allowed by the EOS are shown by the dashed orange and dashed blue lines respectively. {\em Bottom panel:} Percentage of the core volume involved in unsuppressed \durca\ reactions. The shaded region corresponds to the $1\sigma$ blue shaded region in the top panel.}
    \label{fig:MLnopairing}
\end{figure}

The top panel of Figure \ref{fig:MLnopairing} shows the allowed masses for \src, i.e.~the neutron star mass that has the same $L_\nu^\infty$ as \src, as a function of the slope of the symmetry energy $L$. The required mass decreases with $L$, and lies just above the \durca\ threshold mass, shown as a dashed line in Figure \ref{fig:MLnopairing}. At the lowest values of $L$, we find $M \approx 1.8\ M_{\odot}$, whereas for $L\gtrsim 80\ {\rm MeV}$, the mass falls below $1.0\ M_{\odot}$. This result is a consequence of the high efficiency of \durca\ processes which mean that only a small volume of the core is needed to supply the required luminosity.
The bottom panel of Figure \ref{fig:MLnopairing} shows the volume fraction of the core involved in \durca\ reactions, that is, the percentage of core volume above the \durca\ threshold. For stars with the inferred luminosity, the \durca\ volume fraction is around $1$--$4 \%$, similar to the estimate of \cite{Brownetal2018}.

Note that our solutions span a wide range of allowed neutron star masses. A common assumption is that only massive neutron stars can cool by \durca\ reactions, but we see that for the EOS used here the threshold mass is small for high $L$ values. For completeness, we show results for low neutron star masses $\mathrm{M} < 1.0 \, \mathrm{M}_{\odot}$ even though these low masses are unlikely for astrophysical neutron stars (e.g.~see \citealt{Ozel2012} for a discussion of the observed neutron star mass distribution). This implies that non-superfluid cores can explain \src\ only if $L\lesssim 80\ {\rm MeV}$ for the EOSs used here. A similar limit on $L$ comes from the fact that some observed neutron stars are inconsistent with fast cooling (e.g.~\citealt{Wijnands2017}), whereas our non-superfluid models predict that all neutron stars would have \durca\ if $L$ were larger than $80\ {\rm MeV}$. However, these conclusions are relaxed when we include nuclear pairing, as we show in the next section.

\subsection{Effect of superfluidity}\label{sec:SF}

We next include the reduction in neutrino emissivity due to nuclear pairing. By suppressing neutrino emission in regions of the core that are above the \durca\ threshold density, superfluidity can lead to neutrino emission from either a reduced region of the core, or from a shell surrounding the superfluid core. These possibilities are shown schematically in Fig.~\ref{illustration}b and c.

\begin{figure}
\centering
\includegraphics[width=0.47\textwidth]{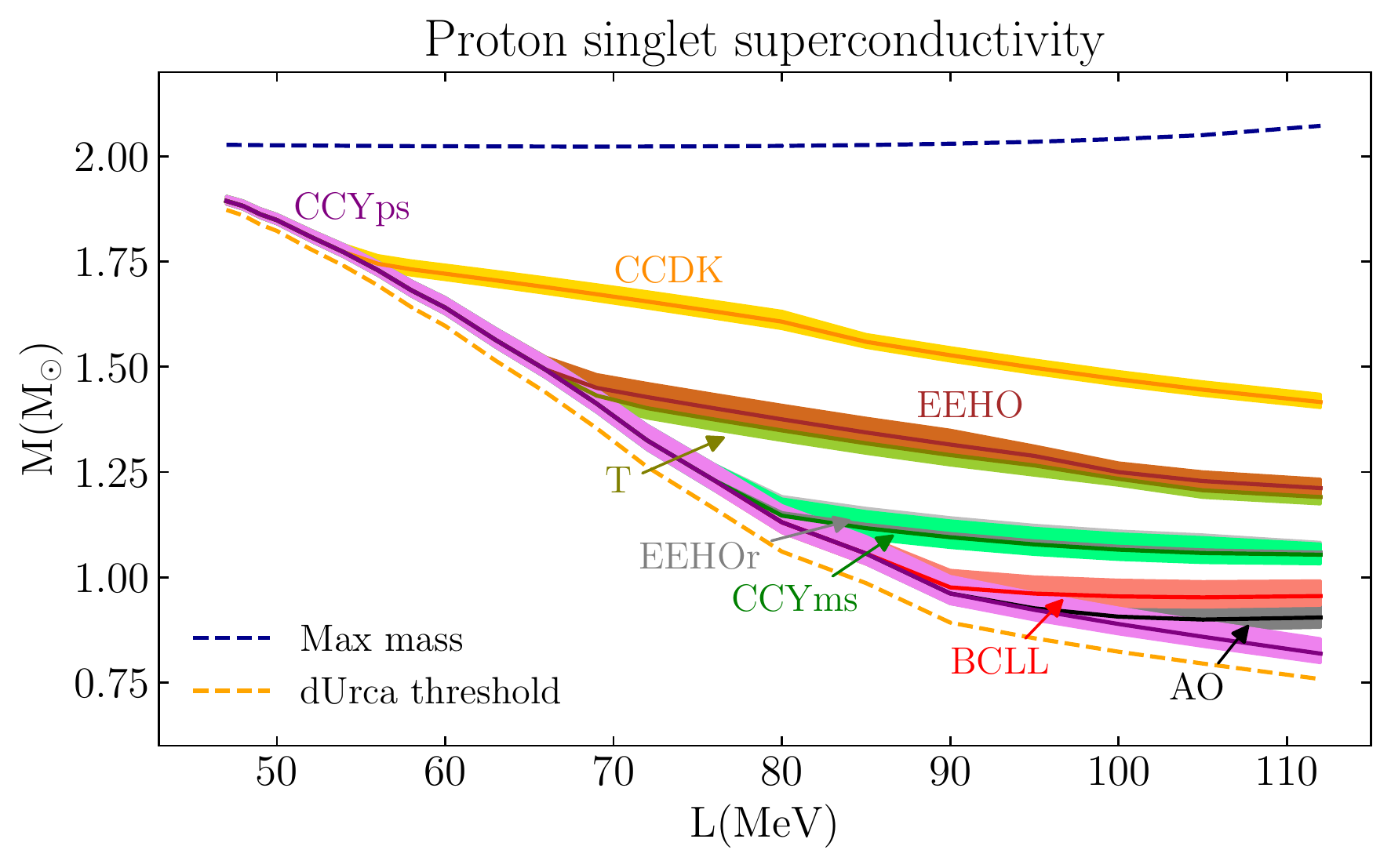}
\includegraphics[width=0.47\textwidth]{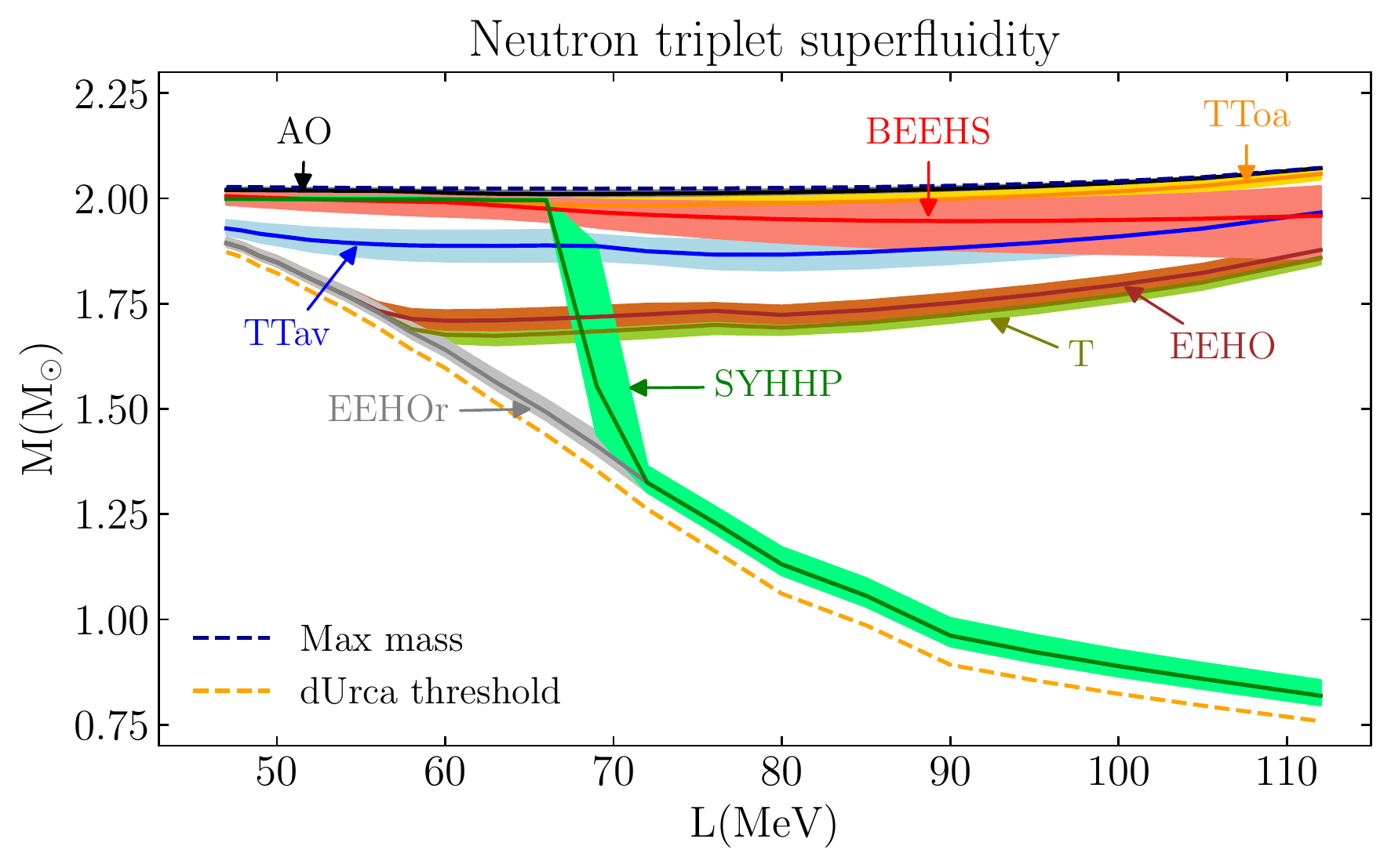}
\caption{As Fig.~\ref{fig:MLnopairing}, but including proton (top panel) and neutron (bottom panel) superfluidity.
In the upper panel, the proton gap models BS and CCYps give the same results as the case with no-pairing shown in Fig.~\ref{fig:MLnopairing}; we show only CCYps (in purple) in the Figure.
In the lower panel, neutron gap model EEHOr (in grey) corresponds to the no-pairing curve shown in Fig.~\ref{fig:MLnopairing}.\label{fig:newR}
}
\end{figure}

We first consider neutron and proton pairing separately to explore the role of each. The results are shown in Figure \ref{fig:newR}, where again we show the allowed neutron star mass as a function of $L$. The effects of nuclear pairing on the allowed masses are substantial, especially for neutron superfluidity. Due to the superfluid suppression of \durca\ emissivity, the effective of onset of \durca\ is delayed to higher density and the mass is increased in most cases (compare Fig.~\ref{fig:MLnopairing} and Fig.~\ref{fig:newR}). In addition, for a given gap model, there is a wider range of inferred masses (a wider color band around the solid lines) than in the no pairing case, because superfluidity smooths the transition to dURCA emission, which means that $L_\nu^\infty$ increases more slowly with increasing $M$ than in the no pairing case. Even though the NT critical temperatures are lower than PS, as shown in Fig. \ref{fig:parametrizations}, neutron superfluidity has a larger effect on the required masses because the opening and closing densities for NT are more likely to occur in the region where \durca\ reactions are allowed. Hence, in calculating $L_\nu^\infty$, the opening and closing densities and the width in density of a gap model is more important than its amplitude (as noted for example in the study of isolated cooling neutron stars by \citealt{BeloinHan:2018}).

\begin{figure}
    \centering
	\includegraphics[width=\columnwidth]{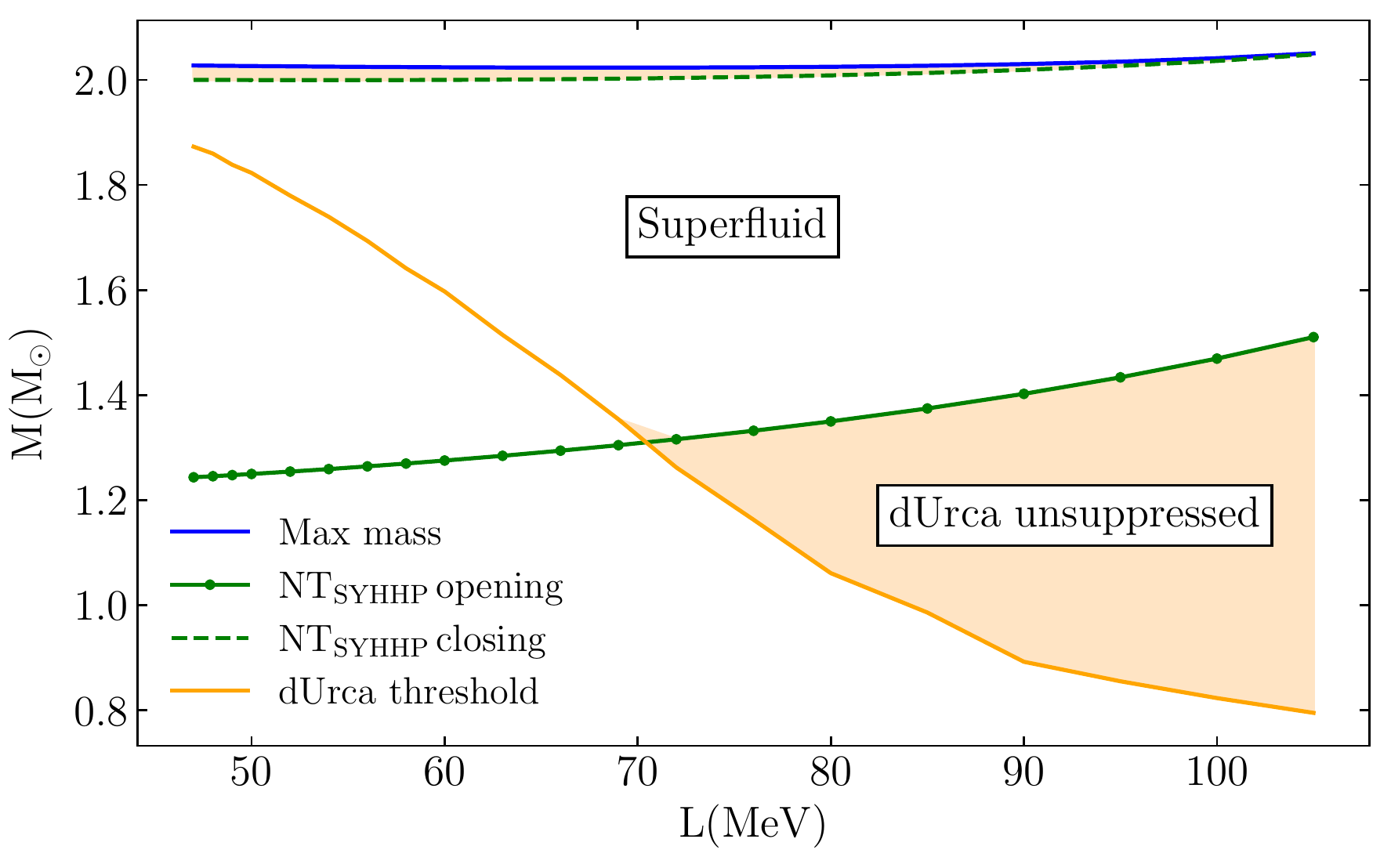}
	\caption{Masses of neutron stars that can cool by \durca\ for the gap model NT SYHHP (shaded regions). \durca\ is unsuppressed at low masses for $L\gtrsim 70\ {\rm MeV}$ or at high masses close to the maximum mass. At intermediate masses, \durca\ is quenched by superfluidity. We show the direct Urca threshold (orange curve), maximum mass (blue curve), and the opening (dotted green) and closing (dashed green) curves of NT SYHHP as a function of $L$.}
	\label{nt_syhhp}
\end{figure}

The ordering of the PS curves in the upper panel of Figure~\ref{fig:newR} follows the ordering of the gap closing density. For $L \leq 52$ MeV ($M \geq 1.75 \, \mathrm{M}_{\odot}$), all the PS gaps close before the central density reaches the \durca\ threshold and the PS gap results are the same as the no pairing results. Two PS gaps, BS and CCYps, close before the onset of \durca\ for all $L$ and hence give the same results as the no-pairing case. The other gaps predict increasing mass as the gap closing density increases: in order of increasing density, these are AO, BCLL, (CCYms,EEHOr), (T,EEHO), and CCDK. The pairs (CCYms, EEHOr) and (T,EEHO) have very similar gap closing densities (see Fig.~\ref{fig:parametrizations}) and therefore give very similar allowed mass ranges. So we see that the role of the PS gap is to delay the effective onset of \durca\ to higher density (from $\rho_\mathrm{\durca}$ to $\rho_{\rm closing}$) and therefore higher masses.

The NT gaps are more complicated because they have different orderings of $\rho_{\rm opening}$, $\rho_{\rm closing}$, relative to $\rho_\mathrm{\durca}$. EEHOr has $\rho_{\rm closing}<\rho_\mathrm{\durca}$ for all $L$ and so gives the same results as the no pairing case. Apart from the gap SYHHP, which we discuss below, the curves again increase in mass following the ordering of the closing densities: (T,EEHO), TTav, BEEHS, TToa, AO, where again we bracket together T and EEHO which have similar closing densities and give similar mass constraints. Because the neutron gaps close at a much higher density than the proton gaps, the NT results for gaps that have $\rho_\mathrm{closing}>\rho_\mathrm{\durca}>\rho_\mathrm{opening}$ give larger neutron star masses than the PS gaps: allowed masses are $\gtrsim 1.65$--$1.8\ M_\odot$, depending on $L$. 

The gap model NY SYHHP is an interesting case. The distinct shape of this curve in Figure \ref{fig:newR} is a direct consequence of the shape of this particular gap, which is narrow and peaks at higher density than the other NT gaps in Figure \ref{fig:parametrizations} (this gap model is a phenomenological model developed to fit the observed cooling of Cas A, see \citealt{Shternin2011}). In Figure~\ref{nt_syhhp}, we show the regions of $M$ and $L$ where \durca\ reactions are allowed somewhere in the neutron star core (orange shaded regions) or are suppressed by superfluidity (unshaded region). At large $L \gtrsim 70$ MeV, the NT SYHHP gap opens after the onset of \durca\ reactions ($\rho_\mathrm{opening}>\rho_\mathrm{\durca}$), leading to a range of masses $\lesssim 1.2$--$1.4\ M_\odot$ which cool by \durca\ without any superfluid suppression. At smaller values of  $L \lesssim 70$ MeV, the core is superfluid already at the densities where \durca\ reactions are allowed. Masses close to the maximum mass, however, have central densities that exceed the closing density of the gap $\rho_0>\rho_\mathrm{closing}$, so \durca\ reactions can then proceed. In Figure \ref{fig:newR}, where we are looking for solutions with a particular value of $L_\nu^\infty$, we see the transition from solutions at high $L$ close to the \durca\ threshold mass to solutions at low $L$ close to the maximum mass. In the first case, the emission is from the core of the star that is not at high enough density to be superfluid; in the second case, the emission is from the core of the star which has a high enough density that the gap has closed. 

We are able to find a solution that matches \src's neutrino luminosity except for one case, the largest $L=112.7$ MeV in our EOS table with gap model NT AO. In this particular case, even a maximum mass star is not able to reproduce the upper value of $L_\nu^\infty$, although it can reproduce the central value. This result holds when we include proton superfluidity and NT AO neutron superfluidity together (\S \ref{PsNt}); therefore the NT AO gap model at very large $L$'s is disfavored. The reason for this is the very broad shape of the NT AO gap, as well as its large amplitude (Fig.~\ref{fig:parametrizations} shows that AO, TToa and BEEHS all have roughly the same width but only AO fails to fit the data). 

The introduction of pairing relaxes the conclusion from the no-pairing models that $L \gtrsim 80 \, \mathrm{MeV}$ requires low neutron star masses that are likely not realizable in nature. Once pairing is included, Figure \ref{fig:newR} shows that the masses are significantly increased for many of the gap models. As mentioned above, the exceptions are the NT gaps EEHOr and SYHHP (which either close before  or open after $\rho_\mathrm{\durca}$, respectively), and the 4 PS gaps BCLL, AO, BS and CCYps (which close before $\rho_\mathrm{\durca}$), which all allow solutions near the \durca\ threshold.

\begin{figure}
\centering
\includegraphics[width=0.9\columnwidth]{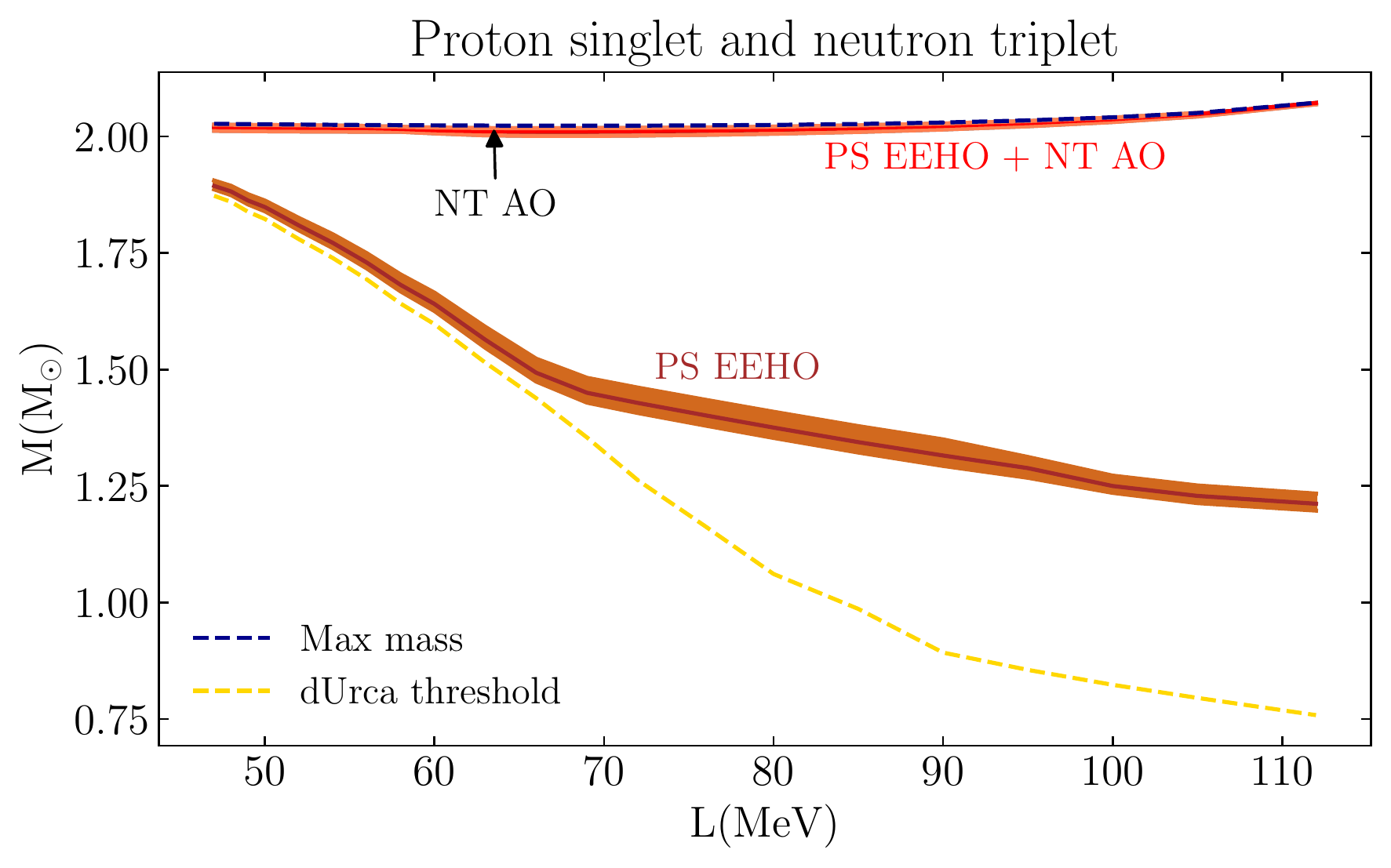}
\includegraphics[width=0.9\columnwidth]{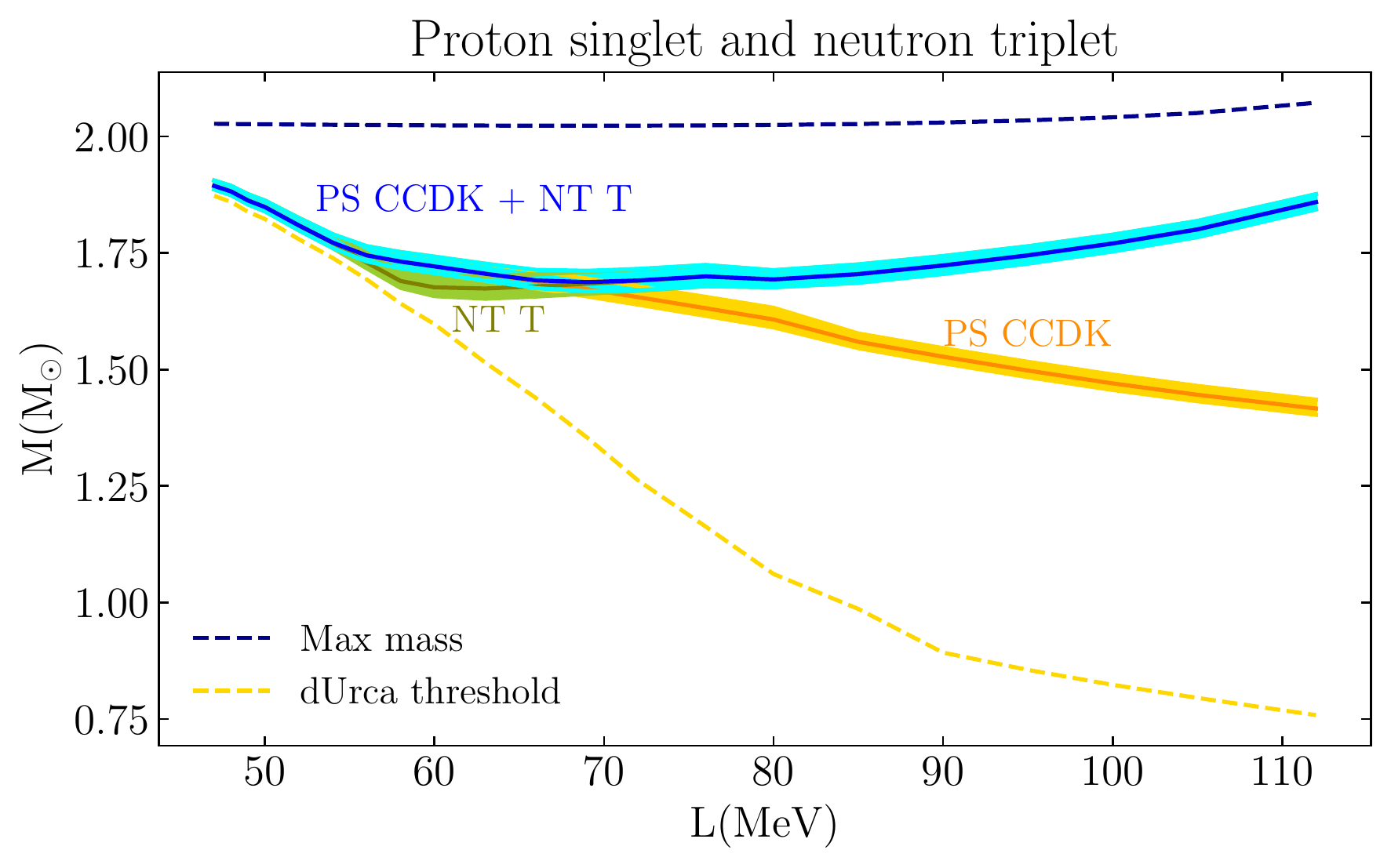}
\includegraphics[width=0.9\columnwidth]{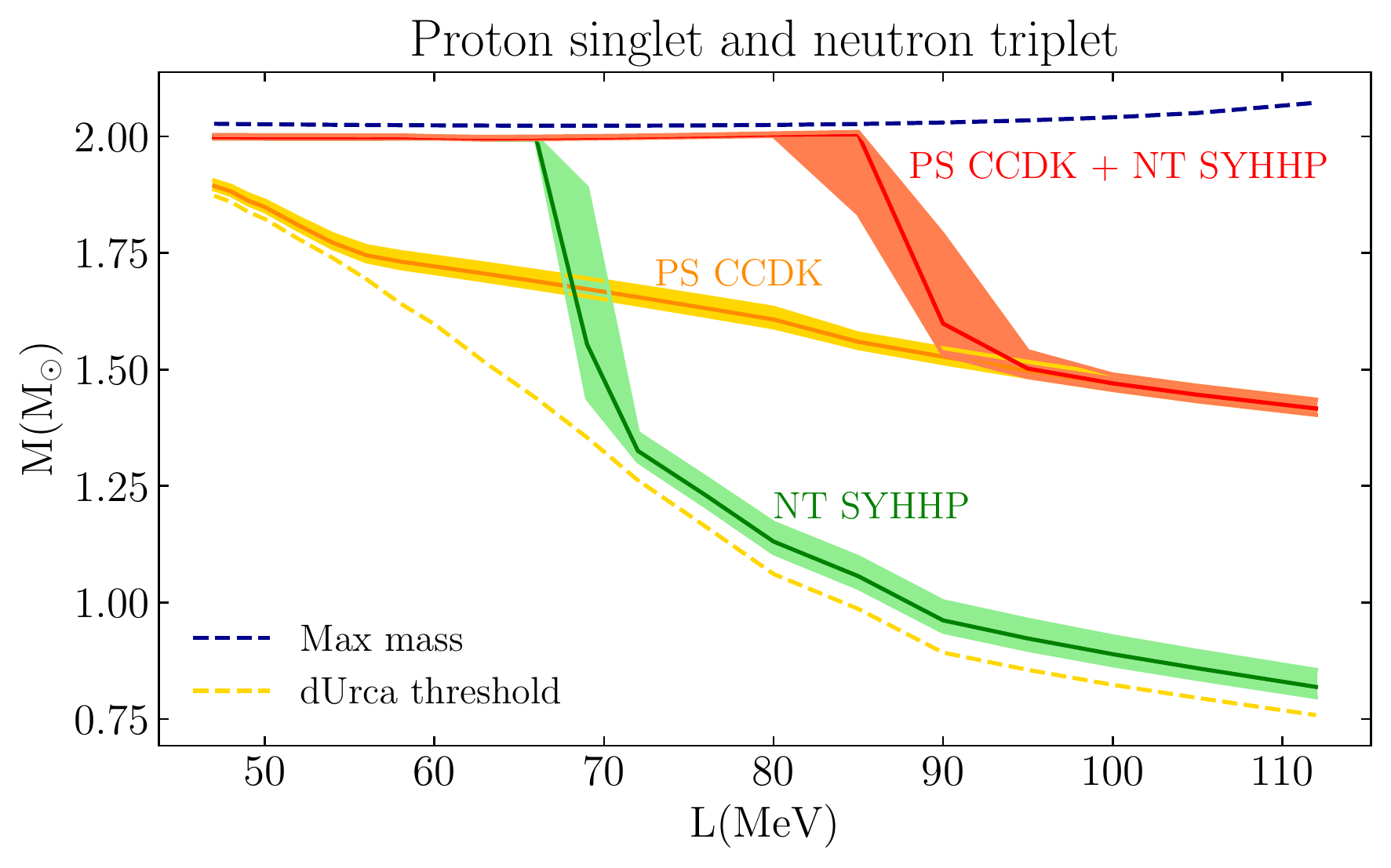}
\caption{As Fig.~\ref{fig:MLnopairing}, but for three examples of NT and PS combinations. In each case, we show the results with proton pairing only, neutron pairing only, and with both neutron and proton pairing included. {\em Top panel}: Gap models $\mathrm{PS} \, \mathrm{EEHO}$ and $\mathrm{NT} \, \mathrm{AO}$. $\mathrm{NT} \, \mathrm{AO}$'s curve is under the curve of the combination. {\em Middle panel}: Gap models $\mathrm{PS} \, \mathrm{CCDK}$ and $\mathrm{NT} \, \mathrm{T}$. {\em Bottom panel}: Gap models $\mathrm{PS} \, \mathrm{CCDK}$ and $\mathrm{NT} \, \mathrm{SYHHP}$.
	}\label{fig:comb1}
\end{figure}

\begin{figure}
\centering
\includegraphics[width=0.48\textwidth]{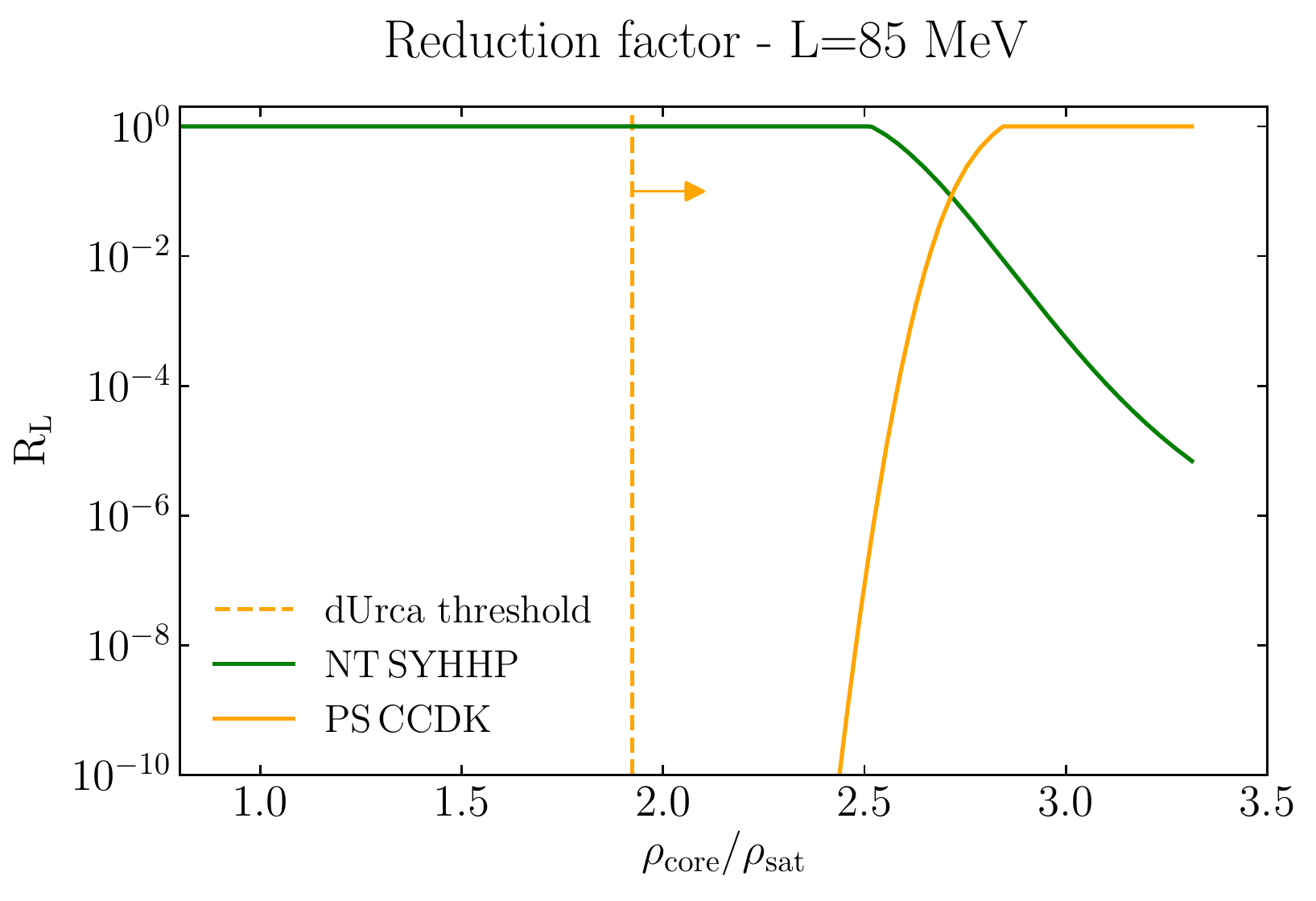}
\caption{Reduction factor $R_L$ of gap models $\mathrm{PS} \, \mathrm{CCDK}$ and $\mathrm{NT} \, \mathrm{SYHHP}$ as a function of core number density over saturation density, for a neutron star with $L = 85$ MeV and $\mathrm{M} = 1.74 \, \mathrm{M}_{\odot}$. The vertical dotted line represents the \durca\ threshold and the region to the right of it, indicated with the arrow, where \durca\ reactions take place.\label{rxn}}
\end{figure}

\subsection{Combination of neutron and proton pairing}
\label{PsNt}

We now include both proton and neutron pairing in the core. Three representative cases are shown in Figure \ref{fig:comb1}. The top panel of Figure \ref{fig:comb1} shows the behavior that we find for most combinations of NT and PS pairings, namely that the neutron superfluid suppression dominates and the effect of proton superconductivity is negligible (a similar conclusion was reached by \citealt{Han2017}). This can be seen in the top panel of Figure \ref{fig:comb1}, where the NT gap model alone produces the same result as the combination of NT and PS (in this case, the solutions are all near the maximum mass for the broad gap model NT AO). The reason that the proton gap does not change the results is that neutron superfluidity is usually active in larger volumes of the core of the neutron star, despite its lower critical temperature, when compared with proton superconductivity gap models. In that case, we obtain the same results as before for the allowed range of inferred masses, $\approx 1$--$5\%$.

There are some pairings of PS and NT gaps, however, for which the choice of the proton pairing gap does change the results. In that case, the results from a model with both PS and NT gaps included can be quite different from those with neutron superfluidity only. Two examples are shown in the middle and bottom panels of Figure \ref{fig:comb1}, both involving the PS CCDK gap which extends to higher density than the other PS gaps (see Fig.~\ref{fig:parametrizations}). In the example in the middle panel, for $L\lesssim 70$ MeV, stars with the inferred luminosity have more of their core volume under $\mathrm{PS} \, \mathrm{CCDK}$ pairing than under $\mathrm{NT} \, \mathrm{T}$ pairing, thus their calculated neutrino luminosity versus mass curve reproduces the previously found $\mathrm{PS} \, \mathrm{CCDK}$ curve. For $L \gtrsim 70$ MeV, $\mathrm{NT} \, \mathrm{T}$ dominates instead and the $L_\nu^\infty$--$M$ curve reproduces the $\mathrm{NT} \, \mathrm{T}$ curve. Note that the width of the calculated mass curve remains narrow, indicating that the range of allowed masses of the star is still small. 

In the lower panel of Figure \ref{fig:comb1}, we show an example in which the solution transitions between NT-dominated high mass solutions (for $L \leq 80 \, \mathrm{MeV})$ to PS-dominated intermediate mass solutions (for $L > 90 \, \mathrm{MeV})$. The transition is significantly shifted in $L$ compared to the NT calculation alone. Note that, at the transition, the range of inferred masses is considerably larger than before, up to $\approx 12\%$ variation in mass. The reason that the results for PS+NT are different from either PS or NT alone is that there are regions in the star where both proton and neutron superfluidity provide comparable suppression factors rather than one or the other dominating. Figure \ref{rxn} shows a specific example of how the reduction factor $R_L$ (see \S \ref{sec:intro_superfluidity}) varies with density for a star with mass $1.74\ M_\odot$ and for $L=85\ {\rm MeV}$ for this choice of gap models. This shows that the proton and neutron reduction rates become comparable at densities higher than the \durca\ threshold, so that both play a role, suppressing emission over a large fraction of the core.

\begin{figure}
\centering
\includegraphics[width=\columnwidth]{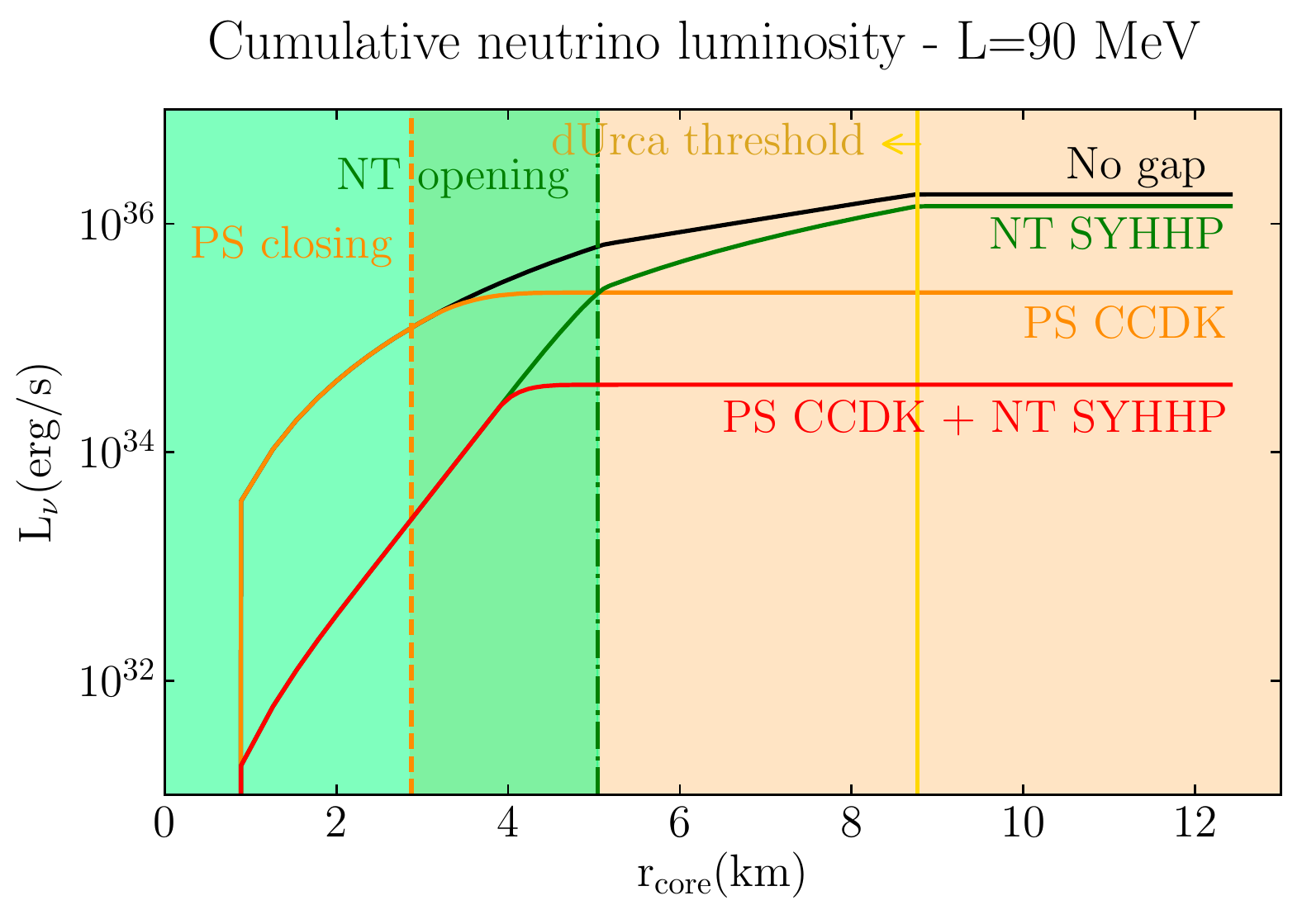}
\caption{Cumulative \durca\ neutrino luminosity as a function of radial distance, for a star with total mass $M = 1.60 \, M_{\odot}$ and $L=90$ MeV. We display the no gap case (black curve) along with the $\mathrm{PS} \, \mathrm{CCDK}$ only case (orange curve), the $\mathrm{NT} \, \mathrm{SYHHP}$ only case (green curve) and the $\mathrm{PS} \, \mathrm{CCDK} \, +$ $\mathrm{NT} \, \mathrm{SYHHP}$ combination (red curve). The region to the left of the vertical yellow line (\durca\ threshold) represents the radii emitting \durca\ neutrinos. The $\mathrm{NT} \, \mathrm{SYHHP}$ opening curve is the dotted dashed green vertical line. The shaded areas are regions under superfluidity, detailed explanation in the text.} \label{fig:lxr}
\end{figure}

To show the different emission regions inside the star in more detail, Figure \ref{fig:lxr} shows the cumulative neutrino luminosity profile for a particular case from the lower panel of Figure \ref{fig:comb1} ($L=90\ {\rm MeV}$ and $1.6\ M_\odot$). The black curve shows the \durca\ luminosity without any superfluidity; the other curves show how this is suppressed as superfluidity is introduced, either NT only, PS only, or NT+PS. The NT+PS curve follows the NT-only curve for the innermost $\approx 4\ {\rm km}$, showing that the NT-pairing suppression dominates there. That region is within the green shaded area on the plot, corresponding to active neutron triplet pairing. At $\approx 5$ km, that gap closes and the proton gap then dominates, represented by the pink shaded area. Its large reduction factor stops the luminosity from accumulating and the curve goes flat, such that the total luminosity is obtained at the innermost part of the core. Note that between $4 \, \mathrm{km}$ and $5 \, \mathrm{km}$ proton reduction rates dominate, even though both nucleon pairings are active. This shifting of neutron and proton superfluidity regions of influence is the signature of transitions as seen in the lower panel of Figure \ref{fig:comb1}.

The fact that in most cases the neutron gap dominates over the proton gap (as in the top panel of Fig.~\ref{fig:comb1}) means that the number of NT and PS gap model combinations that predict low mass stars $(\mathrm{M} \leq 1.0 \, \mathrm{M}_{\odot})$ at large $L$ is actually small. Examples are PS EEHOr+NT SYHHP or PS CCYms+NT EEHOr, which have a late opening of the NT gap or a weak NT superfluidity, respectively. Most of the nuclear pairing combinations investigated favor intermediate to high masses at large $L$. Furthermore, the range of allowed masses is consistently $\approx 5 \%$ for most cases, so that even though superfluidity can change the density range in which significant \durca\ cooling happens, the emitting volume is always a small fraction of the core volume.

\begin{figure}
\centering
    \includegraphics[width=0.48\textwidth]{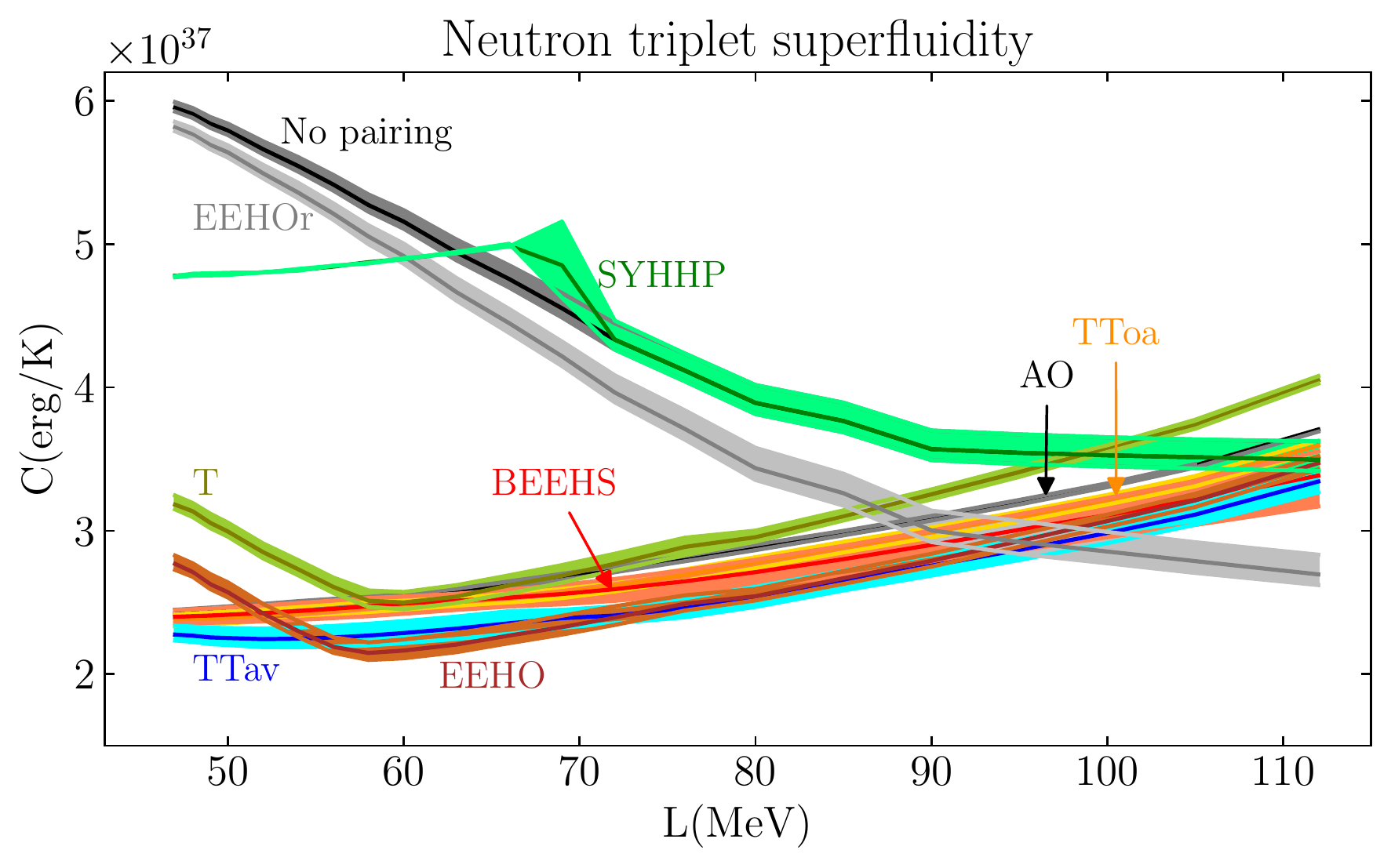}
    \caption{Neutron star heat capacity as a function of $L$ for models that reproduce the inferred neutrino luminosity of \src. The shaded regions around each curve correspond to the $1\sigma$ uncertainty in neutrino luminosity from \citealt{Brownetal2018}. We show results for no pairing and for different NT gap models.\label{heatcap1}}
\end{figure}

\section{The heat capacity of \src}
\label{sec:heatcapacity}

In the previous sections we have shown that a variety of different models can account for the neutrino luminosity of \src. One way to distinguish these different models is through the heat capacity of the neutron star core, which depends on the degree of superfluidity \citep{Brownetal2018}. In this section, we calculate the total heat capacity of our solutions to quantify the nuclear pairing reductions.

In Figure \ref{heatcap1}, we show the total heat capacity of stars that match the neutrino luminosity of \src\ with either no pairing or NT superfluidity only. The value of heat capacity depends primarily on the neutron star mass required to produce the inferred neutrino luminosity (compare each gap model with the corresponding curves in the lower panel of Fig.~\ref{fig:newR}).
In some cases, such as the case with no or weak pairing, where the best fitting mass decreases with increasing $L$, the heat capacity decreases with $L$. In other cases with strong NT superfluid suppression, the allowed masses are larger and tend to increase with $L$; the heat capacity in those cases increases towards larger $L$.

\begin{figure}
\centering
    \includegraphics[width=0.48\textwidth]{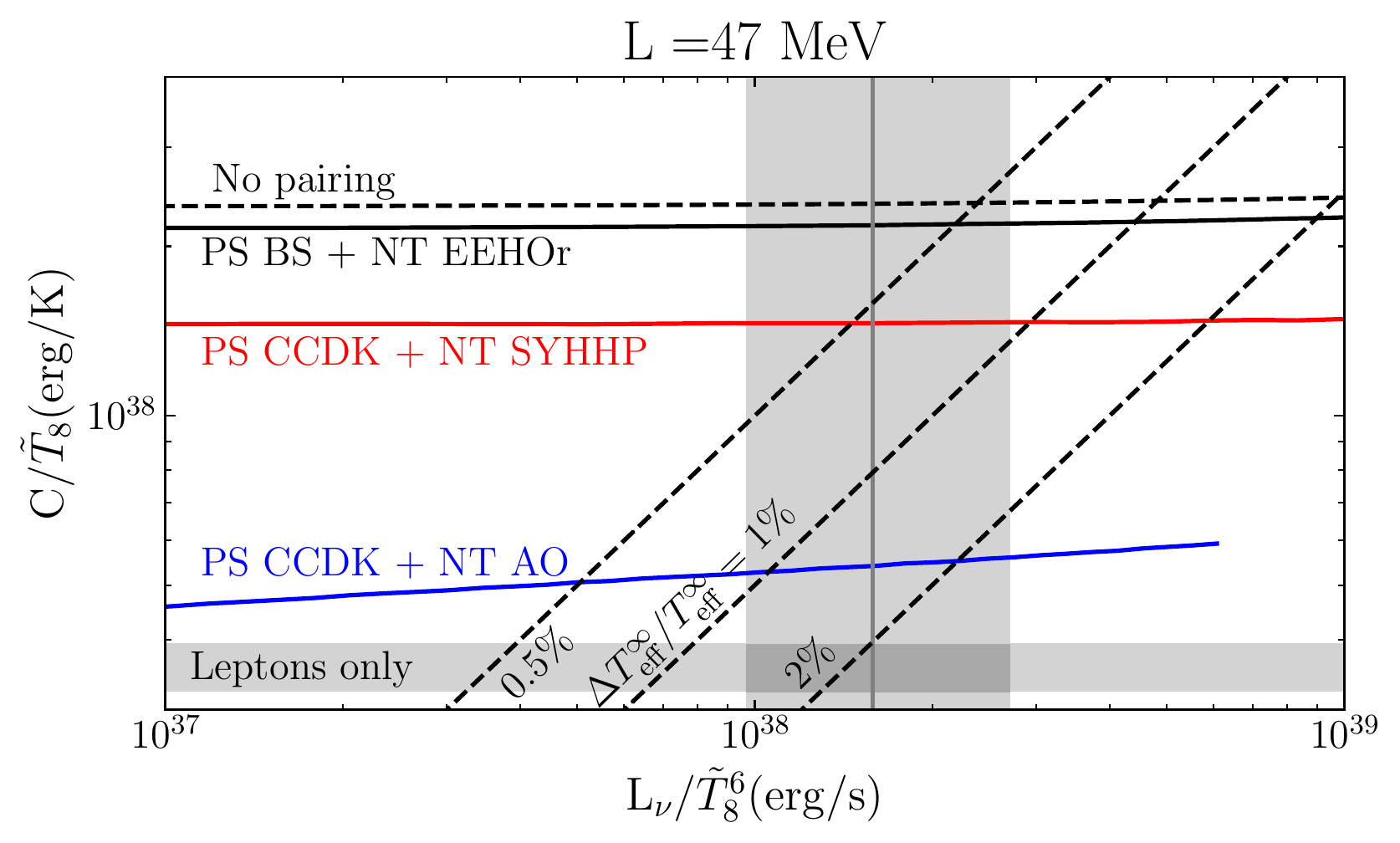}
    \includegraphics[width=0.48\textwidth]{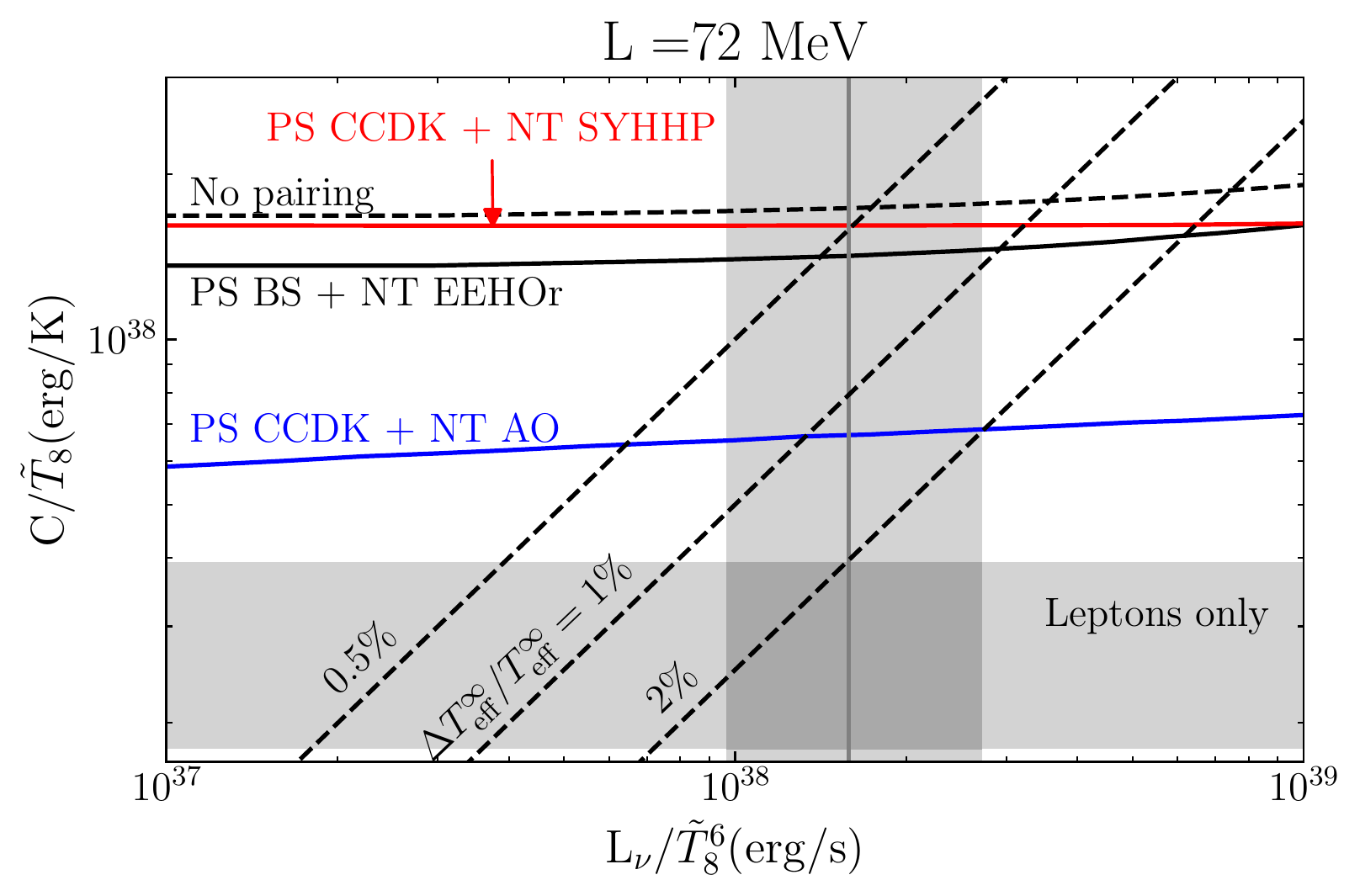}
    \includegraphics[width=0.48\textwidth]{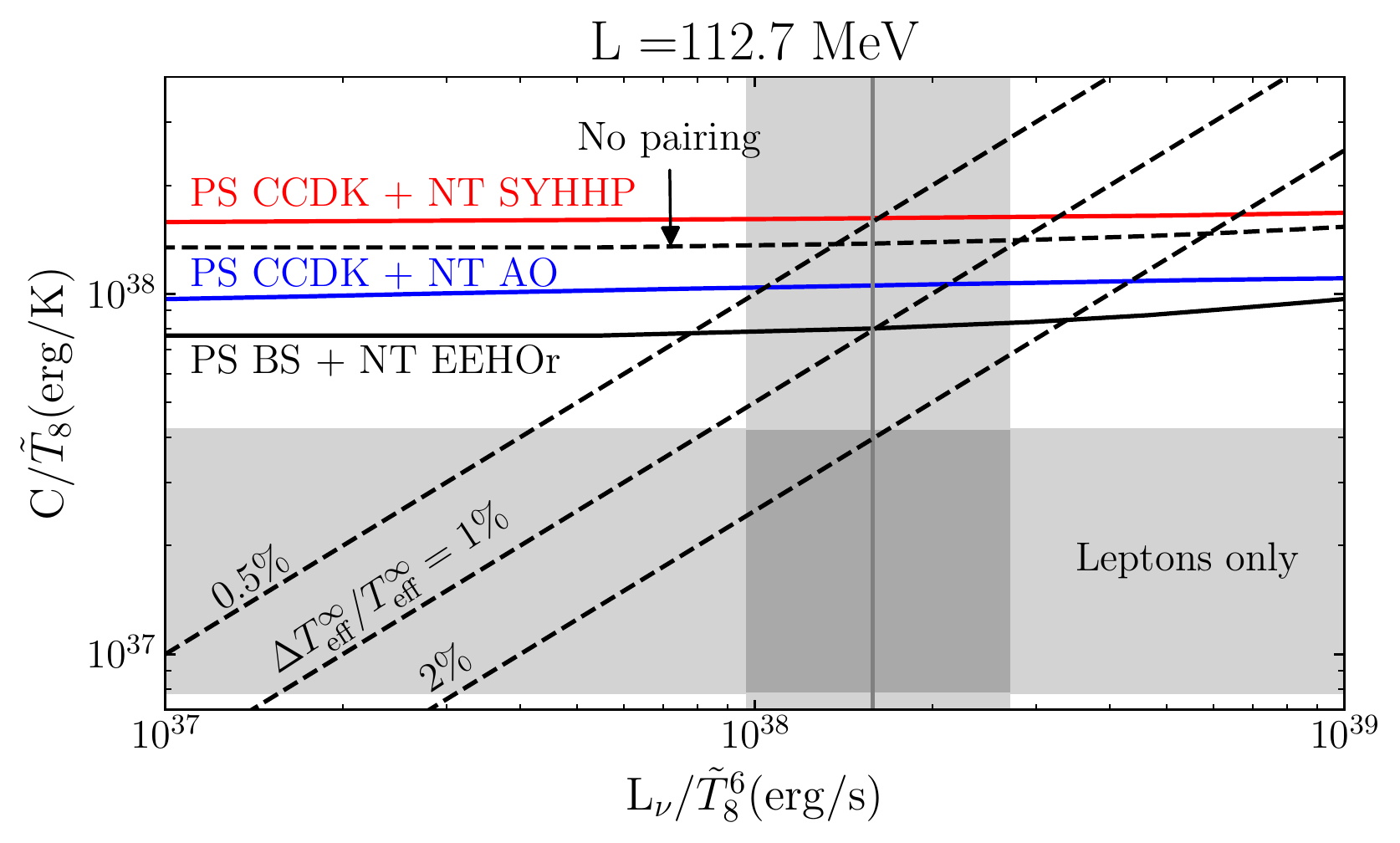}
    \caption{Heat capacity as a function of \durca\ neutrino luminosity for neutron stars with different choices for NT and PS gaps. Models calculated for $L=47$ MeV, $L=72$ MeV and $L=112.7$ MeV, respectively on the top, middle and bottom panels. We show results for no pairing (dashed curve) and for three combinations of NT and PS gaps as indicated (black, red and blue curves). The vertical gray area highlights the inferred neutrino luminosity range for \src\ from \citealt{Brownetal2018}. The horizontal gray area shows the contribution to the heat capacity from leptons. Diagonal dashed lines show the predicted surface temperature variation over a decade given the values of $L_\nu$ and $C$.
    \label{fig:cnu}}
\end{figure}

Figure \ref{fig:cnu} shows the heat capacity as a function of the neutrino luminosity (following \citealt{Brownetal2017} and \citealt{Brownetal2018}) for different combinations of PS and NT gaps.
Unlike neutrino luminosity, for which only the part of the core above the threshold density contributes, the entire volume of the core contributes to the heat capacity. Therefore the superfluid reduction to heat capacity is always visible, even for weak gap models. In addition, the heat capacities of stars with combined superconductivity and superfluidity are significantly smaller than when only one of these pairings is present.  As shown in eqs.~(\ref{reduced3}) and (\ref{reduced4}), the stronger the gap model, that is, the larger its critical temperature compared to the star's temperature, the smaller the star's total heat capacity. The combination of the weakest gap models, $\mathrm{PS} \, \mathrm{BS} + \mathrm{NT} \, \mathrm{EEHOr}$, is the closest heat capacity to the no-pairing case. The combination of the strongest ones, $\mathrm{PS} \, \mathrm{CCDK} + \mathrm{NT} \, \mathrm{AO}$, approaches the value of heat capacity coming from the leptons only (grey band), ie. close to full suppression of the nucleonic contribution to the heat capacity. Other gap models will generate stars with heat capacities between these two limits, as for the example shown of PS CCDK $+$ NT SYHHP. The high efficiency of the \durca\ process together with the small emitting volumes in the case of \src\ lead to the very shallow gradient of the curves in Figure~\ref{fig:cnu}, i.e.~the heat capacity is quite insensitive to $L_\nu^\infty$ since $L_\nu^\infty$ changes rapidly with neutron star mass.

The diagonal lines in Figure \ref{fig:cnu} show the variation in neutron star surface temperature that would be expected over a decade, given the neutrino cooling luminosity $L_\nu$ and the heat capacity $C$. To calculate those curves, we use equation~(24) of \cite{Brownetal2017}, and following their work we assume that the change in core temperature $\tilde{T}$ is related to the change in effective (surface) temperature $T_\mathrm{eff}$ by $\Delta \tilde{T}/\tilde{T} \approx 1.8 \, \Delta T_\mathrm{eff}^{\infty}/T_\mathrm{eff}^{\infty}$.
Our results show that, because the different gap models span almost the full range of heat capacity between the unpaired and lepton-only values, constraints on $\Delta T_\mathrm{eff}^{\infty}/T_\mathrm{eff}^{\infty}$ at the percent level can discriminate between different gap models, clearly indicating whether \src\ is strongly or weakly superfluid. This result, valid for all studied EOS, signals that future measurements of cooling in quiescence of \src\ would constrain nuclear pairing in the neutron star core.

\section{Discussion}
\label{sec:discussion}

\subsection{Cooling of \src\ by \durca}

We find a range of models that reproduce the inferred neutrino luminosity $L_\nu^\infty$ of \src\ with \durca\ neutrino emission. The predicted neutron star mass depends sensitively on the choice of $L$ and gap model (Fig.~\ref{fig:newR}). Since only a small volume of the core undergoing \durca\ reactions is needed to explain the inferred luminosity ($\approx 1$--$4$\% of the core volume), these solutions tend to lie close to either the mass where \durca\ reactions first turn on, i.e.~where the central density first exceeds the \durca\ threshold, or the mass where superfluidity turns off, i.e.~where the central density is large enough that the critical temperature $T_c$ falls below the core temperature, quenching superfluidity and allowing \durca\ reactions to proceed.

For our EOS, the \durca\ threshold mass $M_\mathrm{\durca}$ varies from $\approx 1.85\ M_\odot$ at $L\approx 50\ {\rm MeV}$ to $\approx 1.1\ M_\odot$ at $L\approx 80\ {\rm MeV}$. In this range of $L$ it is possible that the mass of \src\ lies near $M_\mathrm{\durca}$, and superfluidity is not needed. However, for $L\gtrsim 80\ {\rm MeV}$, $M_\mathrm{\durca}$ drops below the smallest mass expected for astrophysical neutron stars, \textit{i.e.} it becomes low enough that all observed neutron stars should be cooling by \durca\ if they are not superfluid. In that case, suppression of \durca\ by superfluidity is essential to allow a solution in which \src\ lies just above the mass where \durca\ cooling turns on (as well as providing a range of lower masses where neutron stars are able to cool by slow neutrino processes).

We find that neutron pairing plays a much more important role than proton pairing in moving the onset of \durca\ cooling to larger masses. This is because the neutron triplet pairing occurs at a higher density than proton singlet (Fig.~\ref{fig:parametrizations}), and so is most likely the cause of superfluidity in the high density regions where \durca\ operates. Proton gap models that close at high density can play a role, e.g.~the CCDK model shown in Figure \ref{fig:parametrizations}, particularly at intermediate values of $L\sim 60$--$80\ {\rm MeV}$. 
In cases where superfluidity moves the onset of \durca\ cooling to higher masses, the predicted mass for \src\ directly reflects the density at which the gap closes (Fig.~\ref{fig:newR}).

We were able to find some combinations of gap models that could not explain \src\ in the case where $L\gtrsim 80\ {\rm MeV}$. Two examples are the combinations PS EEHOr and NT SYHHP or PS CCYms and NT EEHOr. In these cases, the proton gap closes at low density, and the neutron gap is either very weak (in the case of NT EEHOr) or opens at higher density than other models (NT SYHHP), allowing a region of normal matter near the \durca\ threshold, and giving a mass near the threshold mass, ie. $\lesssim 1\ M_\odot$. However, in the majority of cases using other gap combinations we found that superfluidity acts effectively to increase the predicted mass to values above $\approx 1.65$--$1.8\ M_\odot$ depending on $L$. 

In some cases, the superfluid gap extends to high enough density that the predicted mass lies close to the maximum neutron star mass for any value of $L$. Particular examples are the neutron gap models NT AO, TToa, and BEEHS. In one case, with the NT gap AO (the gap with the broadest density range and largest amplitude) and the largest value of $L$ in our EOS table, $L=112.7$ MeV, we were not able to fit the 1$\sigma$ upper limit on $L_\nu^\infty$. Even in other cases where we could fit \src\ adequately with a mass close to the maximum mass, this could cause problems explaining colder sources that require even higher neutrino luminosities, since it does not leave much room to increase the mass and therefore neutrino luminosity further. In particular, the sources \saxj\ and 1H~1905+00 have quiescent luminosities significantly below \src\ (e.g.~see Fig.~3 of \citealt{Potekhin2019}). Indeed, \cite{Potekhin2019} found that a suppressed triplet pairing was necessary to explain \saxj: their standard model of PS BS + NT BEEHS was not able to produce cold enough stars. In this sense \src, with its intermediate quiescent luminosity and small \durca-active volume, is an interesting  data point to add to \saxj\ and 1H~1905+00 when constraining \durca\ emission (e.g.~the study of \citealt{Han2017}), since a combination of gap models that can match \saxj\ for example may not match \src, and vice versa. Note that while \src\ has been included in studies of the population of accreting transients (e.g.~\citealt{Potekhin2019}), the atmosphere is often assumed to have a heavy composition (Fe) which \cite{Brownetal2018} found to be inconsistent with the cooling data.

\subsection{Modelling uncertainties}

A concern for models that explain the neutrino luminosity of \src\ with \durca\ is that the allowed range of neutron star masses is rather small. For the great majority of nuclear pairing combinations, we found a mass range of $\lesssim 5\%$. For some specific EOS and gap model combinations, we can have up to $10\%$ mass change, however, these cases are limited to intermediate $L$ where the dominant pairing transitions from neutron to proton (e.g.~SYHHP gap in Fig.~\ref{fig:newR} lower panel). Given the small number of sources available, it is perhaps unlikely that we would catch \src\ at a time when its mass lies within this small range, although quantifying this probability would require detailed modelling of the evolutionary history.

The range of allowed masses is set by the uncertainty in the derived neutrino luminosity and core temperature.  Based on the modelling of \cite{Brownetal2018}, we have taken a range of about a factor of two in $L_\nu^\infty$. Since the emitting volume is small, this translates to a narrow range of allowed neutron star mass for any given choice of $L$ and gap model. \cite{Brownetal2018} derived the uncertainty in inferred neutrino luminosity of \src\ by marginalizing over the other parameters of their model such as the accretion rate and crust impurity parameter. However, relaxing some of the assumptions of that model would broaden the allowed range of neutrino luminosity. 

The normalization of the accretion rate and the corresponding deep crustal heating rate were included in the marginalization procedure of \cite{Brownetal2018}, accounting for uncertainties in deep crustal heating --- the predicted energy injection ranges from $\approx 0.5$--$2\ {\rm MeV}$ per accreted nucleon in different models \citep{Haensel2008,Fantina2018,Gusakov2021}. However, \cite{Brownetal2018} assumed that the average accretion rate over the last 30 years of observations of \src\ are representative of the longer term average accretion rate (on the timescale to reach thermal equilibrium of the core, hundreds of years for a cold core). Relaxing this assumption would allow for a wider range of accretion rates (e.g.~\citealt{Potekhin2021}). In addition, the marginalization carried out by \cite{Brownetal2018} did not include distance uncertainties, although they estimated that this would change the inferred \durca\ prefactor $L_\nu/\tilde{T}^6$ by less than a factor of two. 

An additional source of uncertainty that we have not included here is in the core temperature. Following \cite{Brownetal2018}, we have taken a fixed value $\tilde{T}=2.5\times 10^7\ {\rm K}$. In fact, for a given measured neutron star surface temperature $T^\infty_{\rm eff}$, the inferred core temperature depends on the envelope model and assumed neutron star mass and radius. The value of core temperature we assume here is for a neutron star surface temperature $T_{\rm eff}^\infty=55\ {\rm eV}$ and a neutron star with mass and radius $M=1.4\ M_\odot$, $R=12\ {\rm km}$ and pure He envelope; including the mass and radius dependence would result in variations of up to $\approx 20$\% in $\tilde{T}$ \citep{Brownetal2017}, or factors of a few in the emitting volume. In addition, the spectral models used to fit the data and obtain the measured $T_{\rm eff}^\infty$ assume fixed values of mass and radius. In addition, although \cite{Brownetal2017} found that an iron envelope could not reproduce the shape of the observed cooling curve for \src, \cite{Potekhin2021} were able to reproduce the cooling with a carbon envelope for a large enough accretion rate, which could be explored further. Although we do not expect it to change our conclusion that the neutron star mass must lie close to the \durca\ onset mass (as allowed by superfluidity), a fully-self consistent study that updates the spectral model and envelope model for each choice of $M$ and $R$ (or $L$) would be worthwhile to quantify the emitting volume more accurately.

\subsection{Alternative fast emission processes}

\begin{figure}
    \centering
    \includegraphics[width=0.48\textwidth]{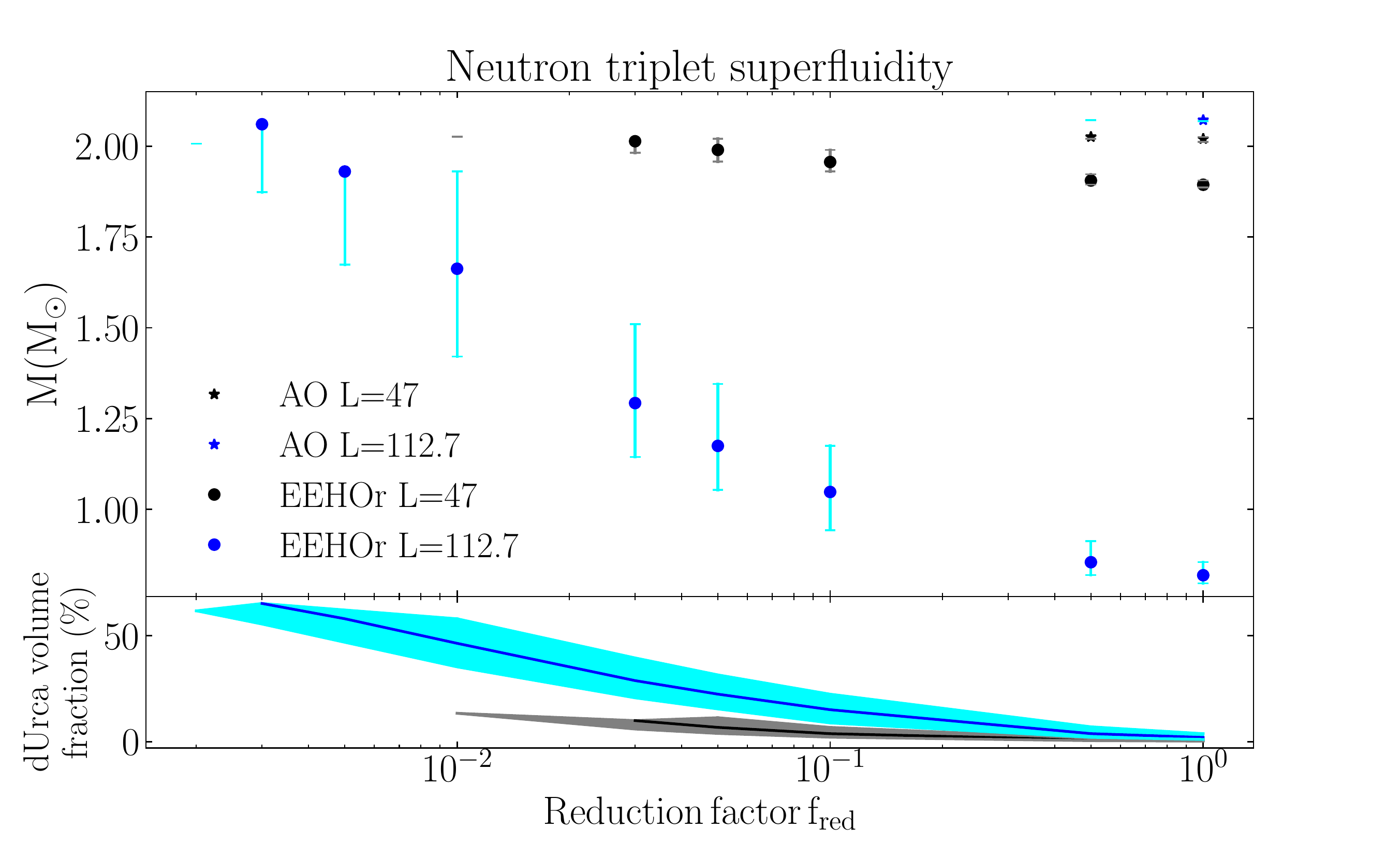}
    \caption{\textit{Top panel:} Neutron star mass that reproduces the neutrino luminosity of \src\, as a function of the \durca\ reduction factor $f_\mathrm{red}$ for neutron star models with neutron pairing only. The error bars correspond to the $1\sigma$ uncertainty in neutrino luminosity from \citealt{Brownetal2018}. We show results for the strongest neutron triplet gap model $\mathrm{NT} \, {\mathrm{AO}}$ (stars) and the weakest neutron triplet gap model $\mathrm{NT} \, {\mathrm{EEHOr}}$ (circles), for $L=47$ MeV (black) and $L=112.7$ MeV (blue), intermediate values of $L$ lie between these curves. \textit{Bottom panel:} Percentage of the core volume involved in dUrca reactions for $\mathrm{NT} \, {\mathrm{EEHOr}}$. The shaded region corresponds to the $1 \sigma$ uncertainty in the top panel. 
    \label{QM_triplet}}
\end{figure}

The small emitting volume for \durca\ provides motivation for considering a less efficient fast process that would result in a larger emission volume and therefore might give a more natural explanation for the observations of \src. Less efficient fast cooling could arise from an uncertainty in the \durca\ prefactor, or from reactions involving other particles such as hyperons, $\Delta$ resonances, or from quark matter. To consistently implement \durca\ cooling from exotic particles would require that we update our EOS to account for the different particle content, and also adjust the \durca\ threshold density accordingly, which is beyond the scope of the current paper. However, as a first check on how our results might change with a less efficient process, Figure \ref{QM_triplet} shows the results of an illustrative calculation in which we keep the same EOS and \durca\ threshold but scale the nucleonic \durca\ emissivity by the constant factor $f_\mathrm{red}$ everywhere in the star. Since a more exotic cooling process likely has a higher threshold density than \durca, the neutrino luminosity we calculate here can be viewed as an approximate upper limit on the emissivity for that case.

The results in Figure \ref{QM_triplet} show that as $f_{\rm red}$ is made smaller, the predicted neutron star mass increases. For example, for the weak superfluid gap EEHOr (which closes before the onset of \durca\ reactions) at $L=112.7\ {\rm MeV}$, the mass rises from $<1\ M_\odot$ (near the \durca\ threshold mass) towards the maximum neutron star mass as $f_{\rm red}$ is reduced to values below $0.01$. This is to be expected since a less efficient process requires a larger emitting volume to generate enough neutrino luminosity. However, our calculations show that, because of redshift corrections and increasing central densities near the maximum mass, the volume fraction increase necessary to reproduce the star's luminosity is not exactly inversely proportional to the reduction factor. This can be seen in Figure~\ref{QM_triplet}, where, for example, for $L=112.7$ MeV, solutions can be found for $f_{\rm red}$ as small as $2\times 10^{-3}$ even though the volume fraction for $f_{\rm red}=1$ is $2$\%.

It is also noticeable that the allowed range of neutron star masses reproducing the inferred luminosity of \src is larger in many cases for $f_{\rm red}<1$, making the model more likely to reproduce the observed \src\ temperature, however this is not always the case. Depending on the nuclear pairing model considered and the star's volume fraction subject to it, the range of masses can also be reduced.

In general, we find that we can reproduce the inferred luminosity of \src\ for any combination of proton and neutron pairing and $L$ as long as $f_\mathrm{red}$ is larger than $\sim 3\times 10^{-3}$\,--\,$3\times 10^{-2}$. In principle this constrains alternative fast neutrino emission mechanisms, e.g.~from pions or kaons which could be suppressed relative to nucleonic \durca\ by a factor of 1000 or more \citep{Yakovlev2001}. This suggests that it would be interesting to further explore models with alternative fast processes that incorporate consistent equations of state and \durca\ thresholds.\newline

\subsection{Future observational and experimental constraints}

Our calculations of the neutron star total heat capacity, combined with its inferred neutrino luminosity, have shown that a future measurement of surface temperature variation in a long time interval could help discriminate between core nuclear pairing models. Figure ~\ref{fig:cnu} shows that our models span the full range of heat capacity, from close to the minimal heat capacity where leptons only contribute, to the larger values where the nucleons in the core are unpaired.
A precise value for that temperature variation, of a few percent, will exclude strong or weak combinations of pairing models, helping to determine the state of matter in the neutron star core. Achieving these observations requires sensitive X-ray observations over many years, and also requires that the source remain in quiescence for this long. \src\ is promising for this, with a mean outburst rate of about 1 every 14 years so far \citep{Maccarone2022}.

We have taken $M$ and $L$ to be free parameters, but our results show that constraints on $L$ and $M$ from future experiments and observations would be extremely constraining for cooling models. For example, if it was shown experimentally that $L\gtrsim 80\ {\rm MeV}$, certain gap model combinations would immediately be ruled out for \src\ in the context of our EOS, i.e.~we need the gap to close at high enough density that the transition to \durca\ is delayed.

The mass of the neutron star in \src\ is currently unconstrained. \cite{Ponti2018} discuss the possibility of measuring the neutron star mass in \src\ using X-ray spectroscopy of the inner regions of the accretion disk. They find that a mass measurement with an uncertainty of about 5\% may be possible with the next generation X-ray telescopes such as Athena (e.g.~\citealt{Nandra2013}). Another possibility is to use spectral fitting of the neutron star thermal spectrum, either in quiescence or during Type I X-ray bursts, to infer constraints on mass and radius, although these methods currently have significant systematic uncertainties \citep{Ozel2016}. The thermal relaxation of the neutron star crust after accretion outbursts also depends on $M$, primarily through its effect on the crust thickness \citep{Brown2009}. In combination with a determination of the neutron star radius, this could lead to tighter constraints on the mass. Comparing the radius range $\approx 11.5$--$13\ {\rm km}$ recently inferred by \cite{Raaijmakers2021} from a variety of astrophysical data, including from NICER \citep{Riley2021}, with Figure~14 of \cite{Brown2009} suggests $M\lesssim 1.6\ M_\odot$ for \src. For the EOS studied here, this would require $L\gtrsim 60\ {\rm MeV}$ and pairing strong enough to delay the onset of \durca\ to this mass (see Figs.~\ref{fig:newR} and \ref{fig:comb1}).

There are various experimental and astrophysical constraints on the value of the slope of the symmetry energy as summarized in \cite{Li:2013ola}. While several experimental results point toward a smaller value of the slope in the range of $40$ to $60$ MeV \citep{Lattimer:2012nd, Drischler:2020hwi}, recent experimental measurements on the neutron skin thickness of $^{208}$Pb \citep{Adhikari:2021phr} implies that $L$ can be much larger, $L = 106 \pm 37$ MeV \citep{Reed:2021nqk}. On the other hand, a very small neutron skin of $^{48}$Ca was measured recently that suggests the $L$-value to be much smaller than all previous constraints combined \citep{Zhang:2022bni}. The lower limit of $L=47$ MeV chosen in this study is a characteristic of the FSUGold2 parametrization below which a self-consistent solution cannot be found. Our exploration suggests $L < 47$ MeV would push the \durca \, threshold even closer to the central density of the maximum-mass neutron star, giving similar results to $L=47$ MeV. While challenging, there are future prospects of a more precise electroweak determination of the neutron
skin at the future Mainz Energy-recovery Superconducting Accelerator (MESA) \citep{Becker:2018ggl} that should allow to constrain $L$ more stringently. From the astrophysical side, the prospects of measuring the radius of a neutron star and its tidal deformability that are both very sensitive to the $L$ value have never been better. NICER aims to measure the neutron star radii with known masses, at a $3\%$ level which should significantly constrain the value of $L$ \citep{Miller:2016kae}. Moreover, future gravitational wave data from binary neutron star mergers should give a strong constraint on the tidal deformability that in turn will constrain $L$ \citep{Fattoyev:2017jql}. 

\section*{Acknowledgements}
We thank David Blascke, Sangyong Jeon, J\'{e}r\^{o}me Margueron and Adriana Raduta for useful discussions and comments, and Ed Brown, Chuck Horowitz, Dany Page, and Sanjay Reddy for many conversations about transient LMXBs and \src\ in particular. This work is supported in part by the Natural Sciences and Engineering Research Council of Canada (NSERC). MM is supported by the Schlumberger Foundation through a Faculty for the Future Fellowship. AC is a member of the Centre de Recherche en Astrophysique du Québec (CRAQ). FF is supported by the Summer Grant from the Office of the Executive Vice President and Provost of Manhattan College. Computations were made on the Beluga supercomputer at McGill University, managed by Calcul Quebec and Compute Canada. 

\software{This work made use of the Python libraries Matplotlib \citep{Hunter2007}, NumPy \citep{harris2020} and SciPy \citep{SciPy2020}.}

\bibliographystyle{aasjournal.bst}
\bibliography{paper-bib}

\end{document}